\documentclass[a4paper,fleqn,usenatbib]{mnras}

\usepackage[T1]{fontenc}
\usepackage{ae,aecompl}

\usepackage{graphicx}	% Including figure files
\usepackage{amsmath}	% Advanced maths commands
\usepackage{amssymb}	% Extra maths symbols
\usepackage{epsfig}
\usepackage{color}
\usepackage{threeparttable}

\newcommand{\ha}{H$\alpha$}
\newcommand{\hb}{H$\beta$}
\newcommand{\redtxt}[1]{\textcolor{red}{#1}}

\title[H$\alpha$ Blobs in MaNGA]{SDSS-IV MaNGA: the physical origin of off-galaxy \ha\ blobs in the local Universe}

\author[X. Ji et al.]{Xihan Ji$^{1,2}$
\thanks{Contact e-mail: \href{mailto:xji243@uky.edu}{xji243@uky.edu}}, 
Cheng Li$^{2}$
\thanks{Contact e-mail: \href{mailto:cli2015@tsinghua.edu.cn}{cli2015@tsinghua.edu.cn}}, 
Renbin Yan$^{3,1}$, 
Houjun Mo$^{4}$, 
Lihwai Lin$^{5}$, 
Hu Zou$^{6}$, 
\newauthor{Jianhui Lian$^{7}$, David V. Stark$^{8}$, Rogemar A. Riffel$^{9,10}$, Hsi-An Pan$^{11}$, Dmitry Bizyaev$^{12,13}$,}
\newauthor{Kevin Bundy$^{14,15}$}
\\
$^{1}$Department of Physics and Astronomy, University of Kentucky, 505 Rose St., Lexington, KY 40506-0057, USA \\
$^{2}$Department of Astronomy, Tsinghua University, Beijing 100084, China\\
$^{3}$Department of Physics, The Chinese University of Hong Kong, Shatin, N.T., Hong Kong S.A.R., China \\
$^{4}$Department of Astronomy, University of Massachusetts, Amherst MA 01003-9305, USA \\
$^{5}$Academia Sinica Institute of Astronomy and Astrophysics, P.O. Box 23-141, Taipei 10617, Taiwan \\
$^{6}$Key Laboratory of Optical Astronomy, National Astronomical Observatories, Chinese Academy of Sciences, Beijing 100012, China \\
$^{7}$Department of Physics and Astronomy, University of Utah, 115 S. 1400 E., Salt Lake City, UT 84112, USA \\
$^{8}$Department of Physics and Astronomy, Haverford College, 370 Lancaster Ave, Haverford, PA 19041, USA \\
$^{9}$Departamento de F\'isica, CCNE, Universidade Federal de Santa Maria,
97105-900, Santa Maria, RS, Brazil \\
$^{10}$Laborat\'orio Interinstitucional de e-Astronomia - LIneA, Rua Gal. Jos\'e
Cristino 77, Rio de Janeiro, RJ - 20921-400, Brazil \\
$^{11}$Max-Planck-Institut f\"ur Astronomie, K\"onigstuhl 17, D-69117 Heidelberg, Germany \\
$^{12}$Apache Point Observatory and New Mexico State University, P.O. Box 59, Sunspot, NM 88349, USA \\
$^{13}$Sternberg Astronomical Institute, Moscow State University, Universitetskij pr. 13, Moscow, Russia \\
$^{14}$Department of Astronomy and Astrophysics, University of California,
1156 High Street, Santa Cruz, CA 95064, USA \\
$^{15}$UCO/Lick Observatory, Department of Astronomy and Astrophysics,
University of California, 1156 High Street, Santa Cruz, CA 95064, USA
}

\date{Accepted XXX. Received YYY; in original form ZZZ}

\pubyear{2021}

\begin{document}
\label{firstpage}
\pagerange{\pageref{firstpage}--\pageref{lastpage}}
\maketitle

\begin{abstract}
\ha\ blobs are off-galaxy emission-line regions with weak or no optical counterparts. They are mostly visible in \ha\ line, appearing as concentrated blobs. Such unusual objects have been rarely observed and studied, and their physical origin is still unclear. We have identified 13 \ha\ blobs in the public data of MaNGA survey, by visually inspecting both the optical images and the spatially resolved maps of \ha\ line for $\sim 4600$ galaxy systems. Among the 13 \ha\ blobs, 2 were reported in previously MaNGA-based studies and 11 are newly discovered. This sample, though still small in size, is by far the largest sample with both deep imaging and integral field spectroscopy. Therefore, 
for the first time 
we are able to perform statistical studies to investigate the physical origin of \ha\ blobs. We examine the physical properties of these \ha\ blobs and their associated galaxies, including their morphology, environments, gas-phase metallicities, kinematics of ionized gas, and ionizing sources. We find that the \ha\ blobs in our sample can be broadly divided into two groups. One is associated with interacting/merging galaxy systems, of which the ionization is dominated by shocks or diffuse ionized gas. It is likely that these \ha\ blobs used to be part of their nearby galaxies, but were stripped away at some point due to tidal interactions. The other group is found in gas-rich systems, appearing as low-metallicity star-forming regions that are visually detached from the main galaxy. These \ha\ blobs could be associated with faint disks, spiral arms, or dwarf galaxies.
\end{abstract}


%%%%%%%%%%%%%%%%%%%%%%%%%%%%%%%%%%%%%%%%%%%%%%%%%%

%%%%%%%%%%%%%%%%% BODY OF PAPER %%%%%%%%%%%%%%%%%%

\section{Introduction}
\label{sec:intro}

Observations of ionized gas in galaxies provide important information 
for our understanding of the evolution history of galaxies. Based on
analyses of the emission line spectra from ionized regions, various 
diagnostics have been used to probe their underlying physical 
properties, such as density structure, ionizing sources, gas-phase metallicity, 
gas kinematics and so on \citep[e.g.][]{1981PASP...93....5B, 1984ASSL..112.....A, 
2001ApJ...556..121K, 2006agna.book.....O, 2019ARA&A..57..511K}. 
Combining these derived properties with spatial locations of the 
ionized clouds in galaxies further provides constraints on the past/recent 
star formation or nuclear activities of galaxies, such as 
inflow/outflow of gas, impact of active galactic nuclei (AGNs), 
and origin of low ionization emission line regions 
\citep[e.g.][]{2012ApJ...747...61Y, 2016MNRAS.461.3111B, 2016ApJ...832..182C, 
2018MNRAS.476.3883L, 2019MNRAS.489..855C}.

Ionized gas also serves as a good tracer of galaxy-galaxy interactions 
in gas-rich systems. Galaxy-galaxy mergers or tidal interactions play 
a fundamental role in galaxy formation and evolution 
\citep[e.g.][]{White-Rees-78, White-Frenk-91, Guo-White-08}. The gaseous
and stellar components inside galaxies could be disrupted and redistributed 
during the process, and the ionization states of the gas can change rapidly 
due to the large amount of released energy. The morphology and properties 
of tidal structures have been vastly studied, but such studies 
are usually limited by the low surface brightness of these tidal features \citep{2008ApJ...689..936J}.
Emission lines from neutral or ionized gas could potentially serve as an alternative to 
unveil the tidal features, in cases where the stellar component of tidal 
features is too weak or overwhelmed by background stellar halo of the central galaxy. Detected ionized gas in the tidal remnants has a variety of origins. It could be ionized by young hot stars formed in the tidally stripped regions. The dense tidal knots might form tidal dwarfs later, exhibiting enhanced star formation and pre-enriched chemical abundances \citep{2000ApJ...542..137H}. AGN
activities triggered by mergers could also play a role in ionizing the gas 
inside the remnants, producing a Hanny's Voorwerp-like object 
(\citealp{2009MNRAS.399..129L,2009A&A...500L..33J}). Other ionization mechanisms, 
including shocks and turbulent mixings, usually exhibit temperature much 
higher than that of photoionization, and are also frequently found in 
merger events \citep[e.g.][]{2011ApJ...734...87R, 2015ApJS..221...28R}.

%The emission lines in tidal remnants could have a photoionization origin either by young OB stars which were formed due to the enhanced star formation during the merger, or by hot evolved stars associated with the remnants. The stripped part might subsequently survive and form so-called tidal dwarfs, exhibiting enhanced star formation activities and pre-enriched chemical abundance (\citealp{2000ApJ...542..137H}). Under certain conditions, AGN activities triggered by mergers could also play a role in ionizing the gas inside the remnants, producing a Hanny's Voorwerp-like object (\citealp{2009MNRAS.399..129L,2009A&A...500L..33J}). Other ionization mechanisms, including shocks and turbulent mixings, usually exhibit temperature much higher than that of photoionization, and are also frequently found in merger events \citep[e.g.][]{2011ApJ...734...87R, 2015ApJS..221...28R}.

Another potential application of observations of ionized gas is the 
identification of faint nearby galaxies. Recent years a group of red 
galaxies called ultra-diffuse galaxies (hereafter UDGs) has been identified 
and studied (\citealp{2015ApJ...798L..45V}). In contrast to their large 
sizes ($\rm R_e = 1.5\sim 4.6~kpc$), their surface brightness is very 
low ($\rm \mu (g,0) = 24\sim 26~mag~arcsec^{-2}$). This population of 
galaxies provide a potential explanation for the so-called 
`missing satellites problem' (\citealp{1999ApJ...522...82K,1999ApJ...524L..19M}).
Recent observations showed that these galaxies have old stellar populations and are rather gas-poor \citep{2018ApJ...859...37G, 2020ApJ...894...32G}.
They could have lost their gas
supply at early times, giving rise to their low stellar mass to dark matter 
mass ratio \citep{2015ApJ...798L..45V}. If the same mechanism that produces the UDGs still works in 
the nearby universe, one might be able to probe the `blue' UDGs, or the 
progenitors of the UDGs that still have ionized gas emission. This will help to test the formation theories of 
UDGs and shed light on the formation rate of this population.

%For optical studies of extra-galactic ionized regions, the observations of \ha\ line play an important role, for reasons as follows. First of all, \ha\ line is one of the strongest optical emission lines. Its intrinsic intensity reflects the number of ionizing photons inside the cloud, and its relative strength to other lines is useful for determining the shape of the ionizing spectral energy distribution (SED) \citep{1981PASP...93....5B}. In addition, in case of H\,{\sc ii} regions, \ha\ line traces the star formation rate (SFR) of the past few Myr \citep{1994ApJ...435...22K, 2012ARA&A..50..531K}. The easily obtainable \ha /\hb\ ratios can be used to estimate the amount of dust extinction through the Balmer decrement \citep{1992ApJ...388..310K}. Moreover, the broadline component of \ha\ line, if observed, is a strong indication of AGN ionization. The spatial locations and shapes of the \ha\ emitting regions also differ for different ionized regions.

%Integral field unit (IFU) surveys are powerful tools to study the spatial distribution of ionized gas inside and around galaxies. 
One interesting outcome from a recent study using the integral field unit (IFU) data from 
the Mapping Nearby Galaxies at Apache Point Observatory survey
(MaNGA; \citealp{Bundy-15, Yan-16b}) is the discovery of a 
mysterious \ha\ blob, which is a large ionized gaseous blob away 
from the galaxy center,  featured by its \ha\ emission. 
\cite{Lin-17} found this \ha\ blob in the galaxy system MaNGA 
1-24145 (z=0.0332, RA=258.84693, DEC=57.43288), which is likely a dry 
merger system with very extended stellar halos. Despite its large 
surface area indicated by the \ha\ emission line, it shows no sign of 
optical counterpart. To explain the formation of this \ha\ blob, \cite{Lin-17} 
and later \cite{Pan-20} proposed the following scenarios.
\begin{enumerate}
	\item It is a gas cloud stripped out from the galaxy due to ram pressure, 
	as the host system falling towards the center of the host cluster of galaxies.
    \item It is a tidal remnant produced by the merger of two galaxies in this system.
    \item It is a UDG, disturbed by the dry-merger system.
	\item It is a transient gas blob expelled and possibly also illuminated by 
	the AGN inside the central galaxy.
    \item It is formed by the cooling of the intragroup medium (IGM).
\end{enumerate}

In order to investigate the physical origin of this \ha\ blob, \cite{Lin-17} used multi-wavelength observations and looked for evidence of underlying structures. They combined the spectroscopic data of this system from MaNGA with their follow-up observation using CFHT/MegaCam and ruled out the first two scenarios, as no stream-like structure or tail-like structure was found. In addition, the morphology of a huge blob seems atypical for a tidal origin. However, even with the deep images from CFHT/MegaCam, they could not find any optical counterpart for the \ha\ blob. It is thus still possible that the present stellar streams together with the optical counterpart of the blob in the system have surface brightness fall below the detection limit of $\rm \sim 26~mag~arcsec^{-2}$. The rest of the scenarios mentioned above cannot be safely ruled out either. %Therefore, \cite{Lin-17} concluded that the origin of this case was still an open question and required more follow up studies.

More recently, \cite{Pan-20} studied this object with new optical, millimeter and X-ray observations \citep{2019MNRAS.488.2925O} and
concluded that the last scenario is most likely, where the observed molecular and ionized gas 
is a direct product of the cooling of the X-ray emitting gas at the location of the \ha\ blob. They found that the ionized and molecular emission in the system can be associated with the part of the hot IGM that is rapidly cooling. Also the estimated cooling time (< 1Gyr) is compatible with the typical values for BCGs hosting warm and cold gas. It seems that the origin of this \ha\ blob is now settled. However, there are still two remaining questions. First of all, is this explanation unique? Due to the complexity of this merging system, other processes could still at play. The available evidence in this single system is compatible with the cooling hot gas scenario, but is not enough to safely exclude all other mechanisms. It is possible that morphologically similar \ha\ blobs in other systems, if found, can have quite different origins. Second, how frequent does this kind of objects form? Can we find more similar objects but at different evolutionary stages? To answer these questions, it is useful to construct a statistical sample. MaNGA, in particular, is an ideal survey for this task.

%The true physical origin of this \ha\ blob, however, still needs more investigations. A statistically important sample would be helpful to further understand the physics behind this kind of object. By looking for other similar systems hosting \ha\ blobs, it is possible to track different evolution stages of these objects. Such a sample would also provide more evidence for/against the potential explanations listed above.

\citet{Bait-Wadadekar-Barway-19} performed the first systematic search of \ha\ blobs and identified a total of 6 \ha\ blobs using the public MaNGA data in SDSS/DR14. They used visual inspection and searched for \ha\ emission with no obvious optical counterpart in SDSS and Legacy Surveys images. They also required the selected \ha\ blobs to have large velocity difference compared to their host galaxies. 
%They were not able to find any underlying structure even in the deep images from the Legacy Surveys. 
About half of their \ha\ blobs and host galaxies show ionization similar to AGNs according to the BPT diagnostics. They therefore concluded that some of these could the faint Hanny's Voorwerp-like objects. Although their study opens a new possibility to investigate \ha\ blobs, their sample is still limited due to their selection criteria.
\cite{Lin-17}'s \ha\ blob does not show large kinematic difference from the host galaxy. Therefore, there could be a considerable number of \ha\ blobs similar to \cite{Lin-17}'s one that have been missed by \cite{Bait-Wadadekar-Barway-19}.
Also, the origin of the \ha\ blobs in their sample that show SF-like ionization is unclear.
%Furthermore, line ratios alone are not a perfect indicator of ionization mechanisms \citep[see e.g.][]{2019ARA&A..57..511K}. 
%There are other information in the spatially resolved MaNGA spectra as well as the optical images not fully explored by \cite{Bait-Wadadekar-Barway-19}, which could potentially provide more clues about the formation of the \ha\ blobs.

With the updated data volume of MaNGA, we attempt to construct a larger statistical sample of \ha\ blobs and reexamine their physical properties. This sample will be obtained through visual inspection with no kinematic cut, in order to make as comprehensive a set of \ha\ blobs as possible. In this work, we make use of the 
MaNGA sample from SDSS/DR15 \citep{Aguado-19} which includes $\sim$4600 unique galaxies, $\sim60\%$ larger
than the sample used in \citet{Bait-Wadadekar-Barway-19}. We identify 
\ha\ blobs by visually comparing the \ha\ map of each galaxy with its optical image,
without attempting to apply selection cuts in any parameters. This careful identification 
process gives rise to a total of 13 \ha\ blobs that are located outside the host galaxies, 
including the one discovered by \cite{Lin-17}. This sample, although still small in size, 
allows us to statistically examine a variety of spatially resolved properties of 
both the blobs and their host galaxies. In addition to the spectroscopic data from MaNGA, 
we also utilize deep imaging data from Beijing-Arizona Sky Survey (BASS; \citealp{2017PASP..129f4101Z}), 
Mayall z-band Legacy Survey (MzLS; \citealp{2016AAS...22831702S}), and Dark Energy 
Camera Legacy Survey (DECaLS; \citealp{2019AJ....157..168D}). We also obtain deep imaging data from the second data release of Hyper Suprime-Cam Strategic Program (HSC-SSP/DR2; \citealp{aihara2019}). 
As we will show, combining these photometric data with the spectroscopic analyses, we are able to improve our understanding of the physical origin of the \ha\ blobs in the local Universe.

The paper is organized as the following. We describe the our data in \S~\ref{sec:data}. 
We introduce our sample selection method and present our sample in \S~\ref{sec:Selection}. 
Analyses about the deep imaging data are shown in \S~\ref{sec:Deep}.
In \S~\ref{sec:photo}, \S~\ref{sec:MZR}, and \S~\ref{sec:kinematics}, we explore potential ionizing sources for \ha\ blobs, their gas-phase metallicities, and their kinematics, respectively. We compare 
our work with previous studies on \ha\ blobs in \S~\ref{sec:discussion}, and summarize 
our findings in \S~\ref{sec:summary}. We also include brief discussions about the \ha\ intensity maps of our sample,
the physical properties of \cite{Bait-Wadadekar-Barway-19}'s sample,
the optical residual images from the Legacy Surveys, and the H\,{\sc i} observations of our sample in Appendix.

Throughout this paper, we assume a $\Lambda$CDM model with H$_0 = 70$\,km\,s$^{-1}$Mpc$^{-1}$, 
$\Omega \rm _m=0.3$ and $\Omega _{\Lambda} =0.7$. We assume a Salpeter initial mass function \citep[IMF,][]{1955ApJ...121..161S} when performing calculations related to stellar masses. All magnitudes are given in the AB system.

\begin{table*}
%	\centering
	\caption{Global properties of the sample galaxies}
	\label{tab:global}
	\begin{tabular}{lcccccccccccc} 
		\hline
		\hline
		No. & MaNGA ID & IAU name  & $z$ &  $NUV-r$ & $\log(\frac{M_\ast}{M_\odot})$ & $R_e$ & $\Delta$RA \mbox{$\dagger$} & $\Delta$DEC \mbox{$\dagger$} & $\rm \Delta v_{H\alpha }$ \mbox{$\dagger$} & $\frac{r_p}{R_e}$ & $\Sigma_{\mbox{H}\alpha}^{\mbox{peak}}$ \\
                    &          &           &   &  (mag) &  & (kpc) & ($^{\prime\prime}$) & ($^{\prime\prime}$) & (km/s) & & ($10^{39}$erg/s/kpc$^2$) \\
		\hline
		1 & 1-339300 & J075437.32+465917.5  & 0.0172 & 3.484 & 9.4 & 1.9 & -5.00 & 8.50 & 39 & 1.78 & 0.78 \\
		2 & 1-52637 & J040723.52-064111.1   & 0.038 & 1.945 & 9.6 & 3.6 & 6.01 & -10.51 & 94 & 2.51 & 0.90 \\
		3 & 1-148597 & J110038.83+501205.1  & 0.0232 & 2.956 & 9.7 & 1.0 & -1.48 & -6.01 & 25 & 2.95 & 0.91 \\
		4 & 1-230052 & J081542.55+255755.2 & 0.0393 & 2.307 & 9.8 & 7.1 & -12.53 & 1.01 & -51 & 1.37 & 2.30 \\
		5 & 1-247975 & J160224.40+424525.4 & 0.0399 & 2.410 & 9.9 & 4.7 & -11.01 & 2.99 & -38 & 1.93 & 0.15 \\
		6 & 1-564242 & J084746.76+540136.0  & 0.0469 & 7.337 & 10.4 & 2.9 & 6.01 & -6.01 & 331 & 2.66 & 0.87 \\
		7 & 1-207914 & J144934.69+525802.2  & 0.0577 & 3.568 & 10.7 & 5.3 & -2.99 & 8.50 & -9 & 1.90 & 1.40 \\
		8 & 1-36779 & J023750.73+003428.6   & 0.063 & 4.464 & 10.8 & 3.7 & 3.49 & -4.50 & 124 & 1.89 & 0.97 \\
		9 & 1-122361 & J080931.34+394806.7 & 0.0639 & 4.553 & 10.8 & 4.2 & 6.48 & -5.51 & -11 & 2.46 & 0.23 \\
		10 & 1-71869 & J075837.71+372859.1  & 0.0401 & 4.981 & 10.9 & 9.4 & 4.00 & 11.02 & 82 & 0.98 & 1.49 \\
		11 & 1-114129 & J213405.08+102518.5  & 0.0774 & 4.595 & 10.9 & 15.6\mbox{***} & -5.51 & -13.00 & 123 & 1.32 & 1.02 \\
		12 & 1-24145\mbox{*} & J171523.26+572558.3 & 0.0322 & 10.554\mbox{**} & 11.1 & 11.4\mbox{***} & -8.06 & 8.68 & -12 & 0.67 & 0.74 \\
		13 & 1-282600 & J122642.32+434704.0  & 0.1115 & 5.409 & 11.5 & 14.1\mbox{***} & -7.99 & -3.49 & 268 & 1.26 & 1.11 \\
		\hline
	\end{tabular}
	\begin{tablenotes}
%	    \centering
        \small
        \item \mbox{*} This is the galaxy system with an \ha\ blob discovered by \cite{Lin-17}.
	    \item \mbox{**} This value is not reliable as the $NUV$ magnitude of this galaxy provided by NSA has an uncertainty comparable to the value.
	    \item \mbox{***} These galaxy systems appear as late-stage mergers and thus the measurement of the effective radius might not be reliable.
	    \item \mbox{$\dagger$} The differences are defined as Parameter(\ha\ blob) $-$ Parameter(galaxy).
    \end{tablenotes}
\end{table*}

\section{Data}
\label{sec:data}

\subsection{Overview of MaNGA data}
\label{sec:MaNGA}

MaNGA is one of the three major experiments of the fourth-generation  Sloan Digital Sky Survey
\citep[SDSS-IV;][]{Blanton-17}.  As the largest IFU survey thus far,
MaNGA aims at obtaining spatially resolved spectroscopy for
$\sim$10,000 nearby galaxies over a 6-year period from July 2014 through June 2020
\citep{Bundy-15,Yan-16b}. MaNGA utilizes 17 hexagonal IFU bundles with five
different field of views ranging from 12$^{\prime \prime}$ to
32$^{\prime \prime}$ to obtain  integral field spectroscopy for target
galaxies \citep{Drory-15}, 12 seven-fiber mini-bundles for flux
calibration, and a set of on-sky fibers for sky subtraction
\citep{Yan-16a}.  Spectra in MaNGA IFU datacubes are taken using the
two dual-channel BOSS spectrographs at the 2.5m Sloan Telescope at the
Apache Point Observatory \citep{Gunn-06, Smee-13}. The spectrographs
cover a wavelength range from 3622\AA\ to 10354\AA\ with a median
spectral resolution of R$\sim2000$.
% reaching a r-band signal-to-noise ratio S/N$=4-8$ per \AA\ per 2$^{\prime\prime}$-fiber at 1-2 effective radius ($\rm R_e$) for a typical exposure time of three hours.

Targets of the MaNGA survey are selected from the NASA-Sloan Atlas 
\footnote{https://www.sdss.org/dr16/manga/manga-target-selection/nsa/}\citep[NSA;][]{Blanton-11},
%which is an SDSS-based galaxy catalog consisting of over 0.6 million galaxies from GALEX, SDSS and 2MASS. 
%By definition, the MaNGA sample has a
%statistically large sample size, 
with an approximately flat distribution in
stellar mass between $\rm 10^9M_\odot$ and $\rm 10^{11}M_\odot$, and a
nearly uniform spatial coverage in units of $\rm R_e$. The sample
includes three subsamples: the Primary and Secondary samples with
spatial coverages out to 1.5$\rm R_e$ and 2.5$\rm R_e$ respectively,
and the Color Enhanced sample to increase the fraction of rare
populations in the color-mass diagram such as low-mass red galaxies
and high-mass blue galaxies. The targets as a whole cover a redshift
range of $0.01$ < z < $0.15$ with a median redshift of $\rm
<z>\sim0.03$ \citep{Wake-17}.
%Details of the MaNGA sample selection can be found in \citet{Wake-17}. 

The raw data of MaNGA are first processed by the Data Reduction Pipeline
\citep[DRP;][]{Law-16}, which produces for each galaxy  a 3D datacube
with a spatial pixel (spaxel) size of $0^{\prime\prime}_.5\times 0^{\prime\prime}_.5$.  
The median PSF of the
datacube has a FWHM of $\sim$ 2$^{\prime \prime}_.$5 \citep{Law-15}.
Therefore, in what follows we require the identified \ha\ blobs to have on-sky sizes larger than the PSF  (or equivalently
5 spaxels), in order for them to be distinguished  from point-like
sources. We also double-check our final sample and make sure that the spatial profiles of these \ha\ blobs are broader than those of unresolved point sources. 
%The observing strategy, survey execution and flux calibration are detailed in \citet{Yan-16a,Yan-16b}. 
The spectral fitting
is done via the MaNGA Data Analysis Pipeline (DAP;
\citealp{2019AJ....158..231W, 2019AJ....158..160B}). This pipeline
utilizes the code {\tt pPXF} (\citealp{2004PASP..116..138C,2017MNRAS.466..798C}) to fit both the stellar continua and emission
lines for galaxy spectra. For each galaxy, DAP first constructs a
best-fitting stellar continuum model for every Voronoi binned spectrum
(\citealp{2003MNRAS.342..345C}), and then fit the individual spectra constituting
each bin based on this model. For each spaxel, the stellar component and the gas component
are fitted simultaneously, with the stellar kinematics fixed by the values taken from the binned spectra. The emission lines are fitted with Gaussian functions. During the fitting of emission lines, DAP requires all lines to have the same velocity, while allowing the velocity dispersions to be different except for the pair of lines in each doublet.

Our sample is drawn from the $\rm 7^{th}$ internal data release of MaNGA, the MaNGA Product Launch 7 (MPL-7), including DRP and DAP products for 4639 unique
galaxies. This sample is identical to MaNGA sample released in SDSS Data Release 15 \citep[DR15;][]{Aguado-19}. Our measurements of \ha\ and other emission lines are all taken from the DAP products.

\subsection{Imaging data}
\label{sec:imaging}

The optical imaging data we use come from large photometric
surveys including SDSS, BASS, MzLS, DECaLS, and HSC-SSP. 

The SDSS imaging data of our galaxies are drawn from Data Release 12
of SDSS (\citealp{2000AJ....120.1579Y}). We use color composite images
generated from images in $g$, $r$ and $i$ bands, of which the median
5$\sigma$ depths are 23.13, 22.70, and 22.20 magnitudes, respectively.

Deeper imaging data of the same galaxies are mainly taken from BASS, MzLS and
DECaLS, which constitute the DESI imaging legacy surveys
(\citealp{2019AJ....157..168D}). BASS uses the 90Prime camera at the
Bok 2.3-m telescope at the Kitt Peak National Observatory (KPNO) to
obtain imaging in SDSS $g$ and $r$ bands, reaching a 5$\sigma$ depth
of approximately 24.0 and 23.5 magnitude \citep{2017PASP..129f4101Z}.
To complement the BASS data, MzLS carries out $z$-band observations in
the same field using the Mosaic-3 camera at the 4-m Mayall telescope
at KPNO, reaching a limiting magnitude of 22.9
\citep{2016AAS...22831702S}. Finally, DECaLS utilizes dark energy
camera mounted at the Blanco 4-m telescope at the Cerro Tololo
Inter-American Observatory (CTIO), obtaining imaging in $g$, $r$ and
$z$ bands down to 5$\sigma$ depths of 24.5, 23.9, and 22.9 magnitude,
respectively. BASS and MzLS cover a total area of 5,000 square
degrees in the northern Galactic cap, while DECaLS covers a total area
of 9,000 square degrees in the southern Galactic cap.

We also use deep imaging data from HSC-SSP/DR2 \citep{aihara2019}. HSC-SSP uses the 8.2-m Subaru Telescope and obtains broad-band $grizy$ images as well as narrow band images for three different layers (Wide, Deep and UltraDeep). The $5\sigma$ depths of the HSC-SSP observations range from $\sim 26$ magnitude for the Wide layer to $\sim 28$ magnitude for the UltraDeep layer \citep{aihara2018}.

Imaging data from the Legacy Surveys and HSC-SSP are much deeper than the SDSS images, making them better options to find the potentially hidden optical counterparts of \ha\ blobs.

\section{Identification of \ha\ blobs}
\label{sec:Selection}

\begin{figure}
	\includegraphics[width=0.45\textwidth]{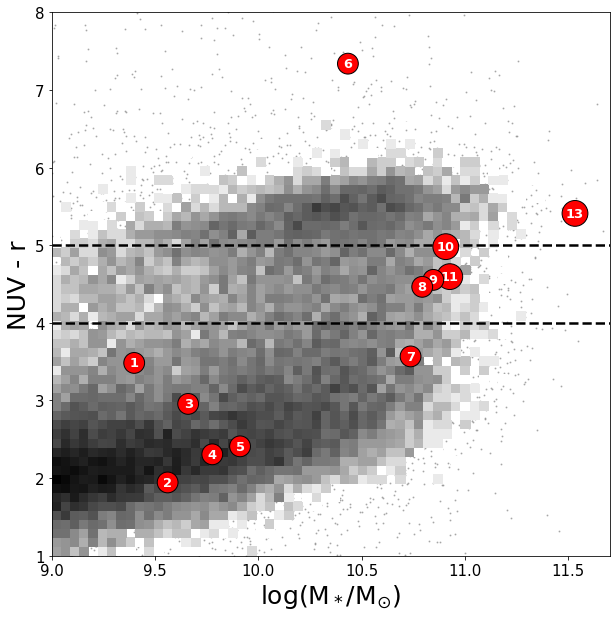}
	\caption{Sample galaxies are plotted as red circles with corresponding serial numbers in the stellar mass versus $NUV - r$ color plane. A representative, volume-limited sample of NSA galaxies are plotted in grey as background. The two dashed lines indicate $NUV - r$ colors of 4 and 5 respectively, commonly used as demarcation lines to separate red-sequence, green-valley and blue-cloud galaxies. For galaxy system {\tt gal12} (MaNGA ID: 1-24145), its $NUV - r$ color is not reliable due to the large measurement error in its $NUV$ magnitude. Thus it is not shown in the figure.}
    \label{fig:mass_color}
\end{figure}

\begin{figure*}
	\includegraphics[width=1\textwidth]{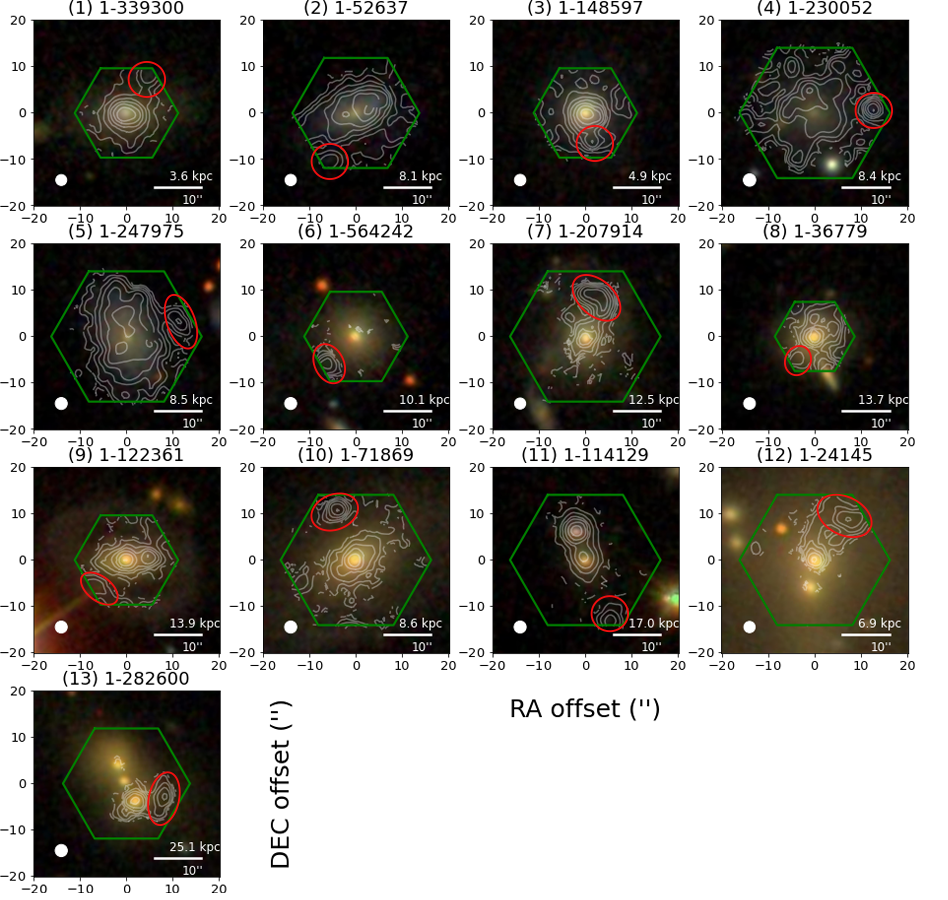}
	\caption{%Color-inverted 
	SDSS $gri$ composite images with contours of \ha\ fluxes. Each set of contours has eight equally spaced levels in logarithmic scale, from the lowest value to the highest value in the system. For system 4, we have masked the foreground star in the south before making the contours. Boundaries of MaNGA's fields of view are shown as green hexagons. For each system, the location of the \ha\ blob is marked with a red ellipse. A scale bar indicating the physical scale of 10 arcsecs is shown for reference, and the FWHM of MaNGA's PSF is indicated by the white circle. The MaNGA ID and serial number is displayed above.}
	\label{fig:sdss_imaging}
\end{figure*}

We identify off-galaxy \ha\ blobs by visually inspecting the
two-dimensional maps of both the flux of the \ha\ line and the $g$-band weighted mean flux for all 4639 galaxies in  the MaNGA DR15 sample. The
detailed procedure is as follows. First, for each galaxy system, we
extract contours from the \ha\ map (which have up to 8 equally spaced levels in logarithmic space from the lowest flux to the highest flux), and overplot them on top of the synthetic $g$-band weighted mean flux images constructed from the MaNGA data. We also apply a signal-to-noise ratio cut which requires the \ha\ flux to have S/N > 3. In this way,  we are able to
identify blobs with \ha\ emission  that are located outside
the host galaxy, as indicated by the synthetic images. Candidates of
off-galaxy \ha\ blobs are  selected to be regions that are covered by
a set of closed contours of \ha\ emission but without apparent optical counterpart in the background. A total of 33 targets are selected by this criterion. We then examine the spectra and the associated quality flags of the spaxels inside the candidate blobs to make sure that they are not foreground stars or
artifacts. In addition, we require that the corresponding spectra show traces of at least one strong optical emission line other than the \ha\ line. This excludes 20 targets, most of which also show spatial profiles close to point sources. As a result, we identify a total of 13 galaxy systems that
have associated \ha\ blobs, including the one discovered by
\citet{Lin-17}. The median S/N of \ha\ line in our \ha\ blobs ranges from 7 to 30. While the median S/N of the $g$-band weighted mean flux is much lower, ranging from 0.6 to 6, with most of the \ha\ blobs having median S/N between 1 and 2.

We note that \cite{Bait-Wadadekar-Barway-19} identified 6 \ha\ blobs based on an earlier sample of MaNGA, SDSS/DR14. Although 6 of our \ha\ blobs are found in DR14, there is only one common source, {\tt hab6} (Plate-IFU: 8724-6102), between the two samples. The poor overlap is a consequence of different selection methods. First of all, \cite{Bait-Wadadekar-Barway-19} required large velocity difference or spatial displacement between \ha\ blobs and host galaxies, while we did not apply similar requirements on these quantities. In addition, we find that different data analysis tools have contributed to the discrepancy, as the \ha\ blobs all have noisy optical continua. Our spectral data is processed by DAP (the public version associated with DR15), while Pipe3D was used in \cite{Bait-Wadadekar-Barway-19}. For three objects in Bait et al.'s sample, this version of DAP masks the spaxels within the \ha\ blobs as they have unreliable kinematic measurements. Finally, the visual identification process is subjective. There is ambiguity in cases where it is unclear whether the \ha\ blobs are simply associated with the disks or other visible structures of the galaxies. This is why the remaining two objects in Bait et al.'s sample are not selected by our method. A more detailed discussion about Bait et al.'s sample can be found in Appendix~\ref{bait}. For consistency and coherence, the main body of this paper will focus on \ha\ blobs identified by our method\footnote{Our conclusion is largely unaffected even if we include the full sample of Bait et al. This is also discussed in Appendix \ref{bait}.}.

\autoref{fig:mass_color} shows the distribution of the host galaxies
of the H$\alpha$ blobs in the $NUV-r$ versus stellar mass
diagram. We assign each galaxy system in our sample with a serial
number from 1 to 13 according to its stellar mass. We will use the following notations hereafter:
for a galaxy with a serial number {\tt x}, we refer to it as {\tt galx}, and the
associated \ha\ blob is called {\tt habx}. When referring to a whole
galaxy system including both the galaxy itself and the \ha\ blob, we
directly use its serial number. For comparison, we have selected a volume-limited sample of
galaxies from the NSA which consists of 26605 galaxies with stellar mass
$M_\ast>10^9M_\odot$ and redshift in the range $0.01 < z < 0.03$. The
distribution of this sample is plotted as the grey-shaded background
in the figure. Measurements of $NUV-r$ and stellar mass are taken
from NSA for both our galaxies and the comparison sample. Our
galaxies can be broadly divided into two subsets: low-mass blue
galaxies ({\tt gal1}, {\tt gal2}, {\tt gal3}, {\tt gal4} and {\tt gal5}) with
$M_\ast<10^{10}M_\odot$ and $NUV-r<4$, and high-mass galaxies with
$M_\ast>10^{10}M_\odot$ and relatively red colors. In the latter
subset, all but one galaxy ({\tt gal7}) have $NUV-r>4$, with four falling in
the green valley regime with $4<NUV-r<5$ ({\tt gal8}, {\tt gal9}, {\tt gal10}, {\tt gal11}), and
two in the red sequence ({\tt gal6} and {\tt gal13}, but we note that {\tt gal13} could also be counted as a green valley galaxy if the dividing lines are made tilted). The galaxy
discovered by \citet[][{\tt gal12} in our sample]{Lin-17} is not shown in
the figure, due to its very large uncertainty in the $NUV$ magnitude
($NUV=-11.0\pm 11.3$) according to NSA. Although the exact value of
$NUV-r$ is uncertain, it should be true that the galaxy is very red, 
based on the quiescent nature of the galaxy as discussed in depth in
\citet{Lin-17}. As a sanity check, we also look into the SFR vs. M$_{*}$ diagram, which leads us to the same classification and confirms that {\tt gal12} lies in the red sequence.
We note that this two-group classification is also largely consistent with the morphological classification. According to the MaNGA morphological catalogue created using the deep learning method by \cite{dominguez2018}, most of the galaxies in the first group are late-type while most of the galaxies in the second group are early-type.

\autoref{fig:sdss_imaging} shows the SDSS $gri$ composite images of our 
sample, overlaid with contours of the \ha\ surface brightness\footnote{The spatial maps of \ha\ intensities can be found in Appendix~\ref{ha_flux}.}. 
The locations of \ha\ blobs are indicated by red ellipses (which do not reflect their actual sizes). The sizes of these \ha\ blobs are different in different systems. If we define the linear size of an \ha\ blob as the size of the contour where the \ha\ flux is half of the value of the blob center, the \ha\ blobs in our sample have sizes ranging from roughly 1.5 kpc to 16 kpc, with 9 of them lying between 3 kpc and 6 kpc. We note that in systems like 1, 2, 6, 9, and 11, the \ha\ blobs seem not fully enclosed by the MaNGA IFUs and their flux contours are truncated. As a result, the actual physical sizes of these blobs could be larger.

\autoref{tab:global} summarizes some
general information of our sample. Measurements of the listed
parameters are taken from the NSA.
We note here a few key features of our sample.
First of all, we have significant detection of \ha\ emission in \ha\ blobs, with
$\Sigma_{\mbox{H}\alpha}^{\mbox{peak}}$ ranging from
$1.5\times10^{38}$erg/s/kpc$^2$ ({\tt hab5}) up to
$2.3\times10^{39}$erg/s/kpc$^2$ ({\tt hab4}) (The values 
are corrected for the dust extinction using the Balmer decrement, assuming 
the intrinsic ratio F(\ha)/F(H$\beta$) = 2.86). 
However, the ionization of these \ha\
blobs are not necessarily dominated by SF activities as their overall \ha\ surface brightness is not very high \citep[see e.g.][]{Zhang-17}. %In a MaNGA-based study of ionized gas regions, \citet{Zhang-17} found that $\Sigma_{\mbox{H}\alpha}$ can be used to separate the two classes of ionized gas regions: star-forming H\,{\sc ii} regions with $\Sigma_{\mbox{H}\alpha}>10^{39}$erg/s/kpc$^2$ and diffused ionized gas regions (DIG) with lower $\Sigma_{\mbox{H}\alpha}$. According to
%\autoref{tab:global}, for the majority of the H$\alpha$ blobs in our sample, 
It is possible that the observed \ha\ fluxes in many of the blobs have significant contribution from the diffuse ionized gas (DIG). 
One could measure the equivalent width of \ha\ line to estimate 
the contamination from non-SF ionization, which we will cover in \S~\ref{sec:photo}.

By selection, the H$\alpha$ blobs are located well away from their nearest 
galaxies. \autoref{tab:global} lists the on-sky position of each \ha\ blob relative to its nearest galaxy.  According to \autoref{tab:global}, the projected separation $r_p$ 
from the center of the optical image is larger than the effective radius of 
the galaxy ($R_e$, defined to be the semi-major axis of the ellipse enclosing half of the total 
light in $r$-band) in most cases. In the most extreme case ({\tt gal3}), the 
blob is found at a distance of $\sim3R_e$. On the other hand, however, 
the contours of the 
%$\Sigma_{\mbox{H}\alpha}$
\ha\ fluxes
continuously cover the 
whole system, coherenty from the host galaxy  all the way to the blob. Also, the relative line-of-sight \ha\ velocities of \ha\ blobs with respect to the galaxy centers are all small except for the system 6 and 13 (a more detailed discussion on kinematics is presented in \S ~\ref{sec:kinematics}).
This implies that the H$\alpha$ emission from the blob may be physically
linked with the H$\alpha$ emission in the host galaxy. We will further 
examine this possibility in the following analyses.

%Here we list the on-sky location of \ha\ blob as well as its
%physical distance to the galaxy center in each  system, given that
%north is up and east is to the left:
%\renewcommand{\labelenumi}{\arabic{enumi}}
%\begin{enumerate}
%\item This \ha\ 
%blob is to the southwest of the central galaxy, and the projected distance
%between their centers is 23.7 kpc.
%\item Northwest, the projected distance is 
%7.8 kpc. 
%\item Southeast, the projected distance is 8.8 kpc. Note that there is 
%a galaxy in the front, to the southwest. 
%\item Southeast, the projected distance 
%is 9.3 kpc. And it looks like the blob is only partly contained in MaNGA's 
%hexagonal view.
%\item Southwest, the projected distance to the nearest galaxy 
%(not the central galaxy) is 15.3 kpc.
%\item North, the projected distance is 
%9.9 kpc.
%\item Northwest, the projected distance is 3.3 kpc. Again, this blob 
%seems only partly enclosed by the data boundary.
%\item Southeast, the projected 
%distance is 8.9 kpc.
%\item South, the projected distance is 3.0 kpc.
%\item West, the projected distance is 8.9 kpc.
%\item North, the projected distance is 9.6 kpc.
%\item Southeast, the projected distance is 12.8 kpc. Also this galaxy system shows 
%a second bright \ha\ core 5.6 kpc to the west.
%\item West, the projected distance 
%is 10.1 kpc. There also seems to be a satellite 6.3 kpc to the east.
%\end{enumerate}

\section{Results}

\subsection{Deep images of the host galaxies}
\label{sec:Deep}

\begin{figure*}
	\includegraphics[width=0.95\textwidth]{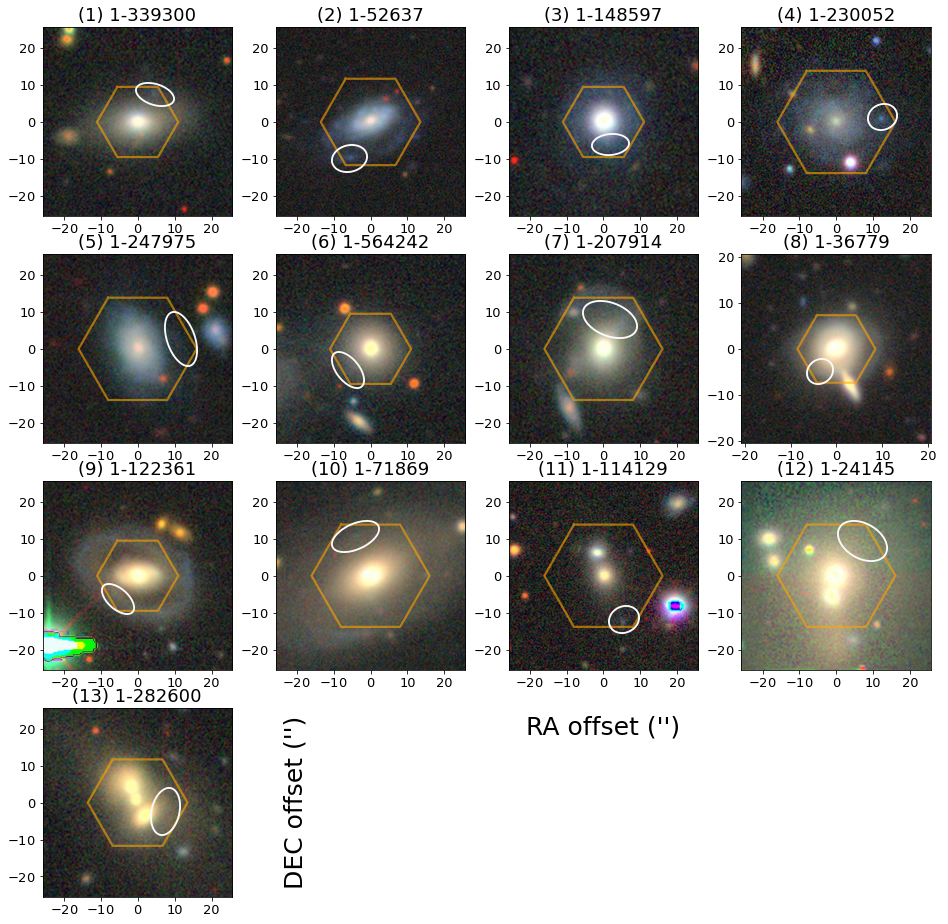}
    \caption{$grz$ color composite images of our sample galaxies from BASS, DECaLS and MzLS surveys.
    For each system, the location of the \ha\ blob is marked with a white ellipse.
    }
    \label{fig:deep_imaging}
\end{figure*}

\begin{figure*}
	\includegraphics[width=0.95\textwidth]{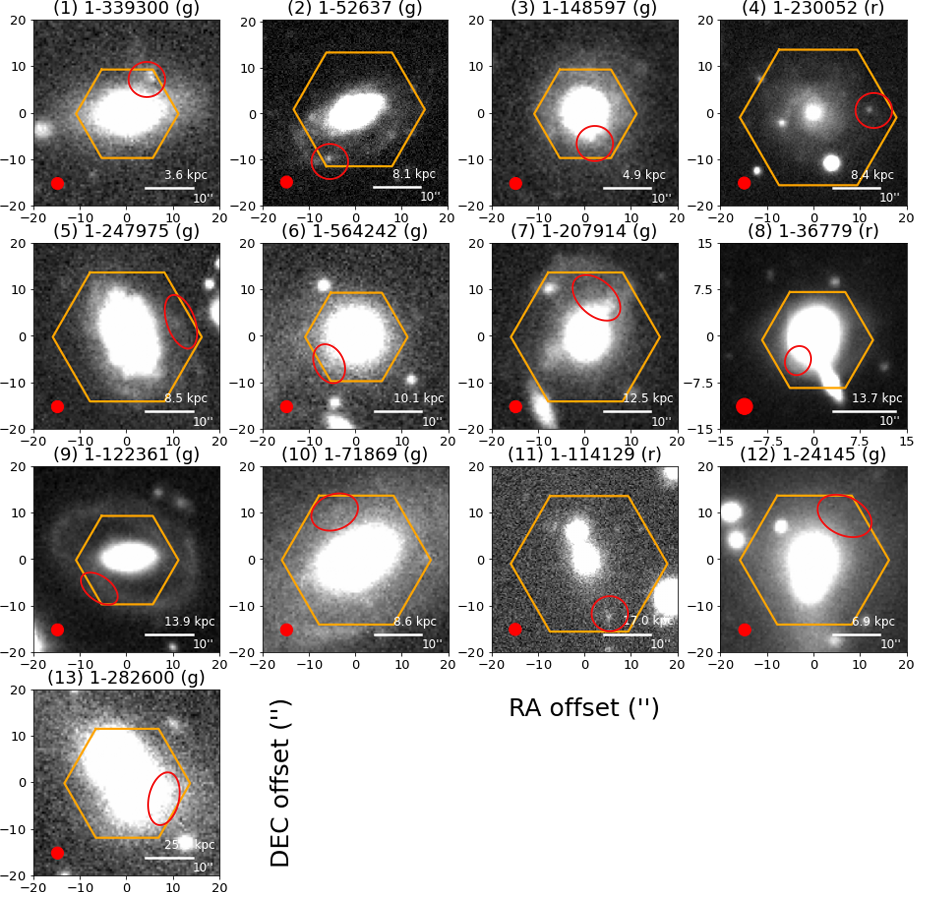}
    \caption{Contrast-enhanced optical images of our sample galaxies from BASS, DECaLS and MzLS surveys. The footprints of MaNGA's fiber bundles are shown as orange hexagons, and the locations of \ha\ blobs are marked with red open elipses. We use $g$-band images for most of the galaxies. For some galaxies, the $r$-band images better reveal the faint optical structures and thus are displayed instead. 
    %In the case of galaxy system 4, only the $z$-band image is available. 
    For each system, the location of the \ha\ blob is marked with a red ellipse. A scale bar indicating the physical scale of 10 arcsecs is shown for reference, and the MaNGA PSF is shown as the red circle.}
    \label{fig:deep_imaging_e}
\end{figure*}

\begin{figure}
    \centering
    \includegraphics[width=0.22\textwidth]{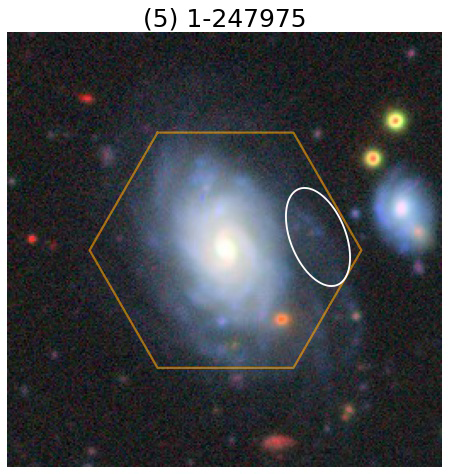}
    \includegraphics[width=0.22\textwidth]{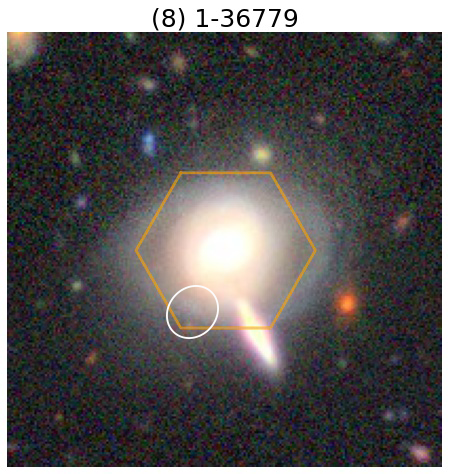}
    \caption{HSC color-composite images of galaxy system 5 (left) and 8 (right). The footprints of MaNGA's fiber bundles are shown as orange hexagons, and the locations of \ha\ blobs are marked with open ellipses.}
    \label{fig:hsc}
\end{figure}

%\textbf{\color{blue} (CL: did we happen to look at the gri composite
%images provided by the MaNGA DRP? The MaNGA data should be much deeper 
%and the SDSS photometric data. Can you make a figure showing the 
%MaNGA images like what you do in \autoref{fig:deep_imaging}?)}

In order to unveil potential stellar components that are associated
with the \ha\ blobs but are too faint to be detected in SDSS images, 
we utilize $g$, $r$ and $z$ band data from BASS, MzLS and
DECaLS, which are much deeper than those from SDSS (see \S~\ref{sec:imaging}).
\autoref{fig:deep_imaging} shows the color-composite images from these surveys. To better reveal the faint optical features, we also utilize contrast-enhanced images\footnote{These images are obtained by adjusting the range of displayed intensities using SAOImage DS9.} in \autoref{fig:deep_imaging_e}. 
We plot the contrast-enhanced images in either $g$, $r$ or $z$ band for the 13 galaxy systems.
In most cases we use the $g$-band image which has the highest signal-to-noise ratio (S/N), while for a few galaxies we opt for the $r$-band in which the faint structures can be more clearly seen.
In addition to the Legacy Surveys images, we also check the HSC-SSP/DR2 images, which are even deeper. Two of our sample galaxies,
{\tt gal5} and {\tt gal8}, have HSC observations. Their color-composite images are shown in \autoref{fig:hsc}.

The 13 galaxy systems in our sample can be categorized in several ways according to their optical images.
To start with, we find a very large fraction of mergers in our sample. Therefore, we can group galaxies based on their environments, i.e. whether they have major companions.
A total of 5 galaxy systems have pairs or trios of (early-type) galaxies: {\tt gal8}, {\tt gal11}, {\tt gal12}, {\tt gal13}, and likely {\tt gal6}. Although {\tt gal6} has no visible companions inside the MaNGA footprint, there is a nearby red galaxy lying to its southeast with a spectroscopic redshift of 0.0489. Their line-of-sight velocity difference is $\sim$ 600 km/s, and could still be interacting. The \ha\ blobs in this group are very likely to be related or under the influence of galaxy-galaxy interactions. 
However, we cannot see clear evidence of common tidal features like streams or shells. These features could have faded and cannot be detected by these surveys. In \autoref{fig:hsc}, we can see {\tt gal8} appear to have an outer ring. It is however not clear whether the ring is associated with galaxy interaction or the perturbation within the galaxy.

Except for system 11, an extended and diffuse stellar halo is present around each galaxy. The stellar halos in these galaxy systems 
seem to reach the locations 
of their H$\alpha$ blobs, but it is non-trivial to tell whether there
are dim underlying optical counterparts that are physically associated 
with these \ha\ blobs. \cite{Lin-17} fitted the envelope of the stellar halo 
surrounding {\tt gal12} using the public code of {\tt GALFIT} and 
concluded that there is no noticeable optical counterpart in the
region of the \ha\ blob. 
We perform a similar analysis by checking the model-subtracted optical images provided by the DESI Legacy Imaging Surveys \citep[see Appendix~\ref{model_sub} for a color-composite figure]{2019AJ....157..168D}. These images were constructed by the {\it legacypipe} pipeline which wraps the source extraction tool {\it The Tractor} \citep{2016ascl.soft04008L}. In {\tt gal6}, {\tt gal12}, and {\tt gal13}, we could not find any residual structure at the locations of the \ha\ blobs that resembles the morphology we see in the \ha\ intensity maps. For {\tt gal11}, the pipeline identified an optical counterpart of the \ha\ blob (which is also visible in the contrast-enhanced image) and subtracted it from the final image. Its $r$-band morphology resembles a small clump or a spot (but we note that its \ha\ flux profile is broader than an unresolved point source). This optical counterpart might indicate a dwarf galaxy or the progenitor of a tidal dwarf, i.e. the dense knot produced by the tidal interaction and emerges from the rest part of the tidal structure that is too faint to be observed. 
For {\tt gal8}, the \ha\ blob seems to reside on the faint outer ring of the central galaxy.
The remaining 8 galaxies all appear to live in isolated systems.
While {\tt gal1}, {\tt gal5}, and {\tt gal7} have
close companions which are located outside the MaNGA IFU, the redshifts of the companion galaxies are 
larger than those of the central galaxies. The companion of 
{\tt gal5} has a spectroscopic redshift of $0.07170\pm0.00007$ \citep{2012ApJ...750..168K}, and the companions of {\tt gal1} and {\tt gal7} have photometric redshifts of $0.242\pm 0.054$ and $0.176\pm0.031$, respectively \citep{2016MNRAS.460.1371B}.
But there is still possibility for them to be involved in interaction with minor companions.

Based on the shapes of the emerging faint optical structures in deep images, we can also categorize our sample galaxies into four subsets. {\tt gal1}, {\tt gal4}, and {\tt gal11} show small clumps or spots at the locations of their \ha\ blobs; {\tt gal2}, {\tt gal5}, and {\tt gal7} show faint spiral arm or tail at the locations of their \ha\ blobs; {\tt gal3} and {\tt gal10} show extended disks or stellar halos; {\tt gal8} and {\tt gal9} show faint outer rings.
Due to their low surface brightness, these features are missing or barely visible in \autoref{fig:sdss_imaging}, but can be seen in \autoref{fig:deep_imaging}, \autoref{fig:deep_imaging_e}, and \autoref{fig:hsc}. The deepest images provided by HSC-SSP best reveal the previously hidden structures. From \autoref{fig:hsc}, it is clear that the \ha\ blob in {\tt gal5} is associated with a faint spiral arm. While the \ha\ blob in {\tt gal8} could be part of the outer ring, similar to the case of {\tt gal9} despite much fainter. 

In the first subset of our sample, the \ha\ blobs have spot-like optical counterparts, which have small physical sizes. While the $g-r$ color of {\tt hab1} is similar to that of its host galaxy (with $g-r \approx 0.6$),
{\tt hab4} and {\tt hab11} are much bluer compared to their host galaxies ($[g-r] _{\tt hab4}\approx 0.2\approx [g-r] _{\tt gal4}-0.2$, $[g-r] _{\tt hab11}\approx 0.7\approx [g-r] _{\tt gal11}-0.2$). The $r-i$ colors of these \ha\ blobs are all greater than 0. Judging from the colors, {\tt hab4} could fall into the category of luminous compact galaxies, but not the extreme cases like green pea galaxies \citep{cardamone2009, izotov2011}. The origin of these \ha\ blobs could be either in situ or tidal, which requires further evidence to tell and is discussed in \S~\ref{sec:discussion}. For the second subset, \ha\ blobs reside in the spiral arms and could simply be SF regions at the outskirt of galaxies. {\tt gal7} is a special case where it is hard to tell whether the single arm in the north is a spiral arm or a tidal arm arced toward the center. Interestingly, its \ha\ blob (indicated by the contours in \autoref{fig:sdss_imaging}) traces the region interior to the arm rather than the visible arm itself. In the third subset, we see very extended disks or stellar halos emerge as the images become deeper. In this case, it is not clear whether the \ha\ blobs are isolated components or part of the disk/halos and more evidence is needed. For the fourth subset, there are outer rings at the locations of \ha\ blobs in both merging and isolated systems. Although \ha\ blobs are likely residing on the rings, whether their formation is in situ or involving galaxy interaction is unknown. Finally, the remaining galaxies ({\tt gal6}, {\tt gal12}, and {\tt gal13}) do not show new structures in deep images.

We conclude that the a large fraction of our \ha\ blobs appear to be in interacting/merging galaxy systems, either in pairs/trios of early-type galaxies ({\tt gal6}, {\tt gal8}, {\tt gal11}, {\tt gal12}, {\tt gal13}) or in systems with potential tidal signatures ({\tt gal7}, {\tt gal8}, {\tt gal9}), although more accurate identification of the tidal features are needed in these cases. These galaxies are all massive red galaxies (the second group defined in \S~\ref{sec:Selection}).
Therefore, the 
deep images presented above prefer
a picture involving galaxy interactions, supporting scenario (ii) and scenario (iii) 
we listed in \S~\ref{sec:intro}. 
There are also \ha\ blobs potentially related to dwarf galaxies or SF regions on spiral arms ({\tt hab1}, {\tt hab2}, {\tt hab4}, {\tt hab5}, {\tt hab7}, and {\tt hab11}), which open a new possibility. They are also compatible with scenario (iii), i.e. being very diffuse galaxies. We note that these candidates are mostly in low-mass blue galaxies (the first group defined in \S~\ref{sec:Selection}).
We exclude scenario (i) as only {\tt gal12} is found near a group center. However, we cannot 
completely exclude scenario (iv), as AGN activities could also be 
related to mergers. We will explore the potential contribution 
of AGNs in \S~\ref{sec:photo}. As a closing remark of this section, 
we have found that the morphology and environments of \ha\ blobs 
in our sample have both similarities and differences, which indicate 
that these \ha\ blobs we see are not necessarily the same class of 
object observed by \cite{Lin-17} or \cite{Bait-Wadadekar-Barway-19}. Alternatively, it is 
possible that they are related by some common physical processes 
(e.g. galaxy interactions), but differ in the exact conditions that 
determine their morphology and spatial locations. Therefore, we 
need more information about their properties in order to understand 
how (or whether) they can be put into a consistent physical picture. 
In the following sections, we perform spectral analyses and investigate 
the relationship between \ha\ blobs and their host galaxy systems 
based on their ionizing sources, the derived chemical abundance, and kinematics of ionized gas.

\subsection{Ionizing source}
\label{sec:photo}

It is critical to understand 
the ionizing mechanisms behind these \ha\ blobs, as it would help us understand 
the recent activities of these galaxy systems. In addition, most of the metallicity estimators rely on a clear classification of the
ionizing source. Optical diagnostic diagrams are powerful tools to perform 
this task, especially in our case where only a few strong optical emission 
lines are available \citep{1981PASP...93....5B, 1987ApJS...63..295V}. 
We examine the distribution of our sample galaxies and blobs in both the 
[N\,{\sc ii}]- and [S\,{\sc ii}]-based BPT diagrams, using the demarcation 
lines introduce by \cite{2001ApJ...556..121K}, \cite{2003MNRAS.346.1055K}, and
\cite{2006MNRAS.372..961K}. The results are shown in \autoref{fig:2BPT}. 
It is clear that the spaxels inside the \ha\ blobs are mostly classified as 
star-forming (SF) regions, while the spaxels of the host galaxies are a mix 
of SF regions and low-ionization (nuclear) emission line regions, or LI(N)ERs 
in short \citep{1980A&A....87..152H}.

\begin{figure*}
    \includegraphics[width=0.95\textwidth]{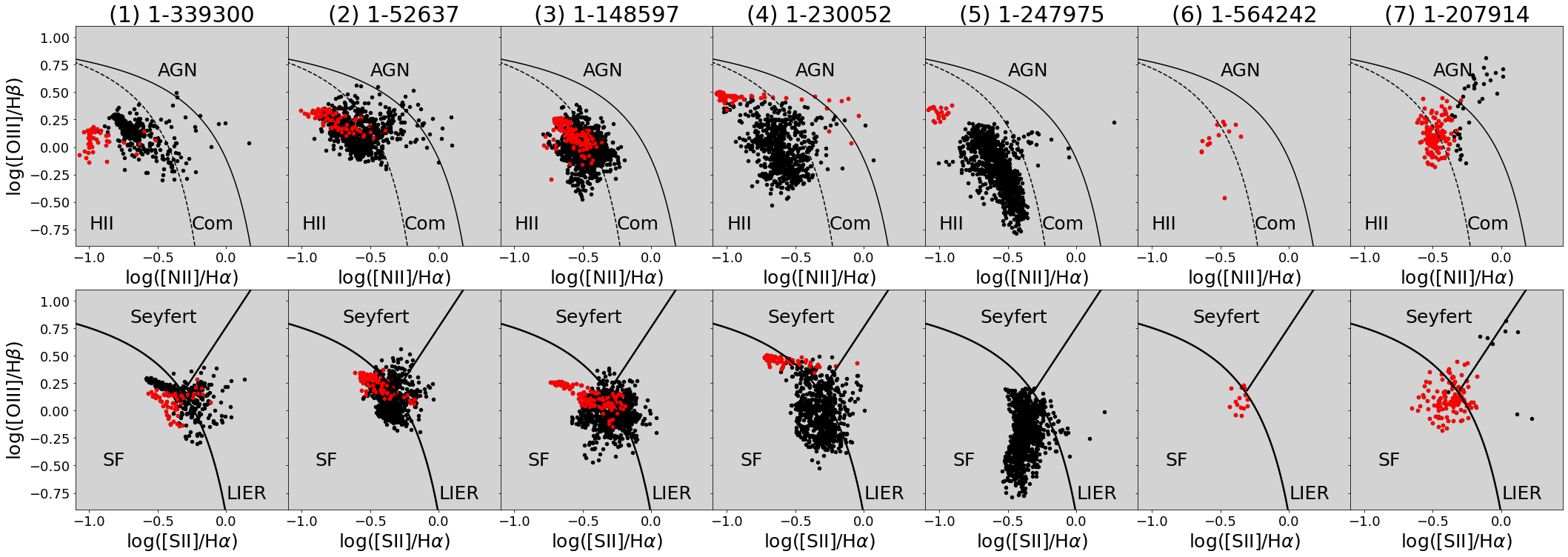}
\end{figure*}

\begin{figure*}
    \includegraphics[width=0.95\textwidth]{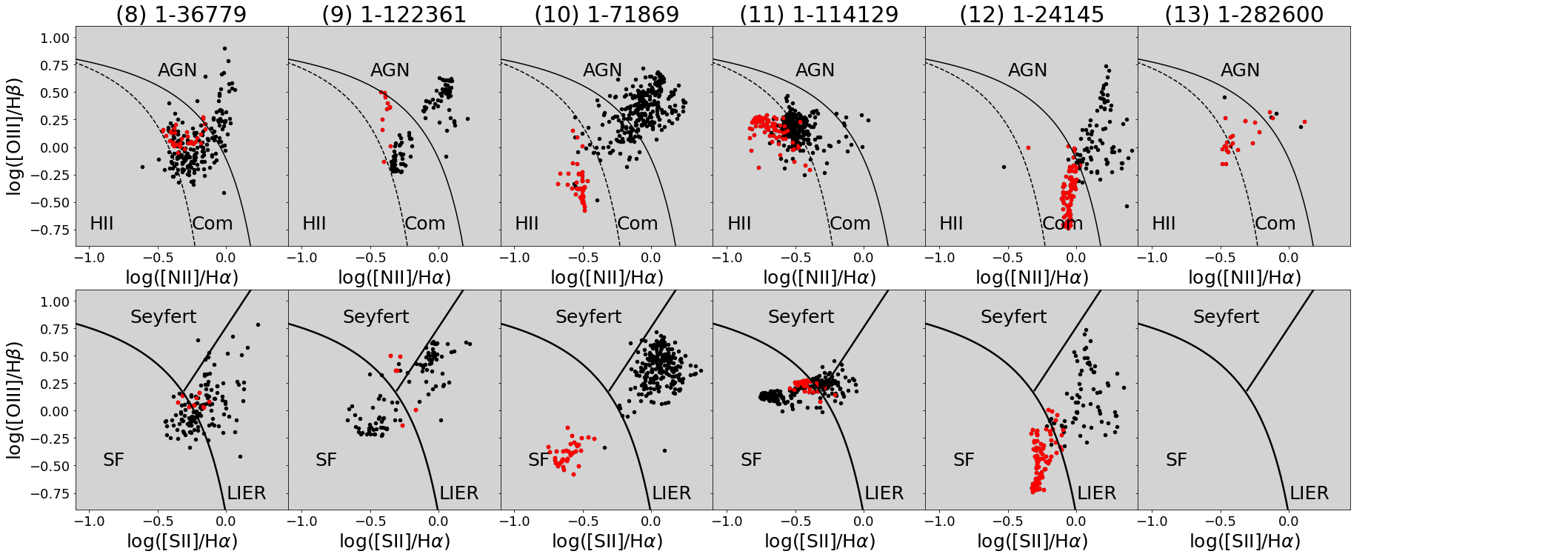}
    \caption{Distribution of our sample spaxels in [N\,{\sc ii}] and [S\,{\sc ii}] BPT diagrams. Red points correspond to spaxels inside \ha\ blobs, and black points correspond to spaxels belong to the host galaxies.}
    \label{fig:2BPT}
\end{figure*}

Looking at the distribution of our sample spaxels in BPT diagrams, we can 
visually identify two different groups. Galaxy system 1, 2, 3, 4 and 5 
are similar in the sense that the spaxels are roughly following the typical 
`SF sequence', which are often seen in large samples of H\,{\sc ii} regions 
or SF galaxies \citep[e.g.][]{2003MNRAS.346.1055K}. The \ha\ blobs in these 
systems tend to occupy the upper left locus of the data distribution, 
indicative of relatively low gas-phase metallicity and relatively high ionization 
parameter. The other group includes galaxy system 7, 8, 9, 10 and 12, which present a continuous distribution from the SF region to the AGN 
region, with the \ha\ blobs sitting in the lower left position of the data 
distribution. On the other hand, these \ha\ blob spaxels locate to the lower 
right position relative to the SF sequence, indicating overall high metallicities, 
which is consistent with the fact that their host galaxies are all high-mass 
galaxies (see \autoref{fig:mass_color}). Comparing this with the result we obtain in \S~\ref{sec:Selection} and \S~\ref{sec:Deep}, one can see that these two groups also show certain differences in color, morphology, and environments, which serves as another piece of evidence 
that the visually found \ha\ blobs in our sample might have different origins. 
The \ha\ blobs in the former group (consisting of galaxy system 1, 2, 3, 4 and 5) 
appear as low-metallicity and off-galaxy starburst or SF regions, while the \ha\ 
blobs in the latter group (consisting of galaxy system 7, 8, 9, 10 and 12) seem 
to be associated with galaxy-galaxy interactions/mergers in which the main galaxies are already quenched.

Back in \S~\ref{sec:Deep}, we found {\tt hab1}, {\tt hab4} and {\tt hab11} have spot-like optical counterparts. These counterparts have blue colors and they are indeed populated by SF regions. It is interesting to know whether these \ha\ blobs are blue compact dwarfs undergoing intense star formation. Besides the evidence from their optical colors, we checked the equivalent widths of \hb\ and [O {\sc iii}]$\lambda 5007$. {\tt hab4} has the highest equivalent width, with the peak EW(\hb )$\approx 60$ \AA\ and peak EW([O {\sc iii}])$\approx 230$ \AA. Although its equivalent widths are not as extreme as green pea galaxies (of which EW(\hb )$\gtrsim 100$ \AA ), it falls into the sequence of luminous compact galaxies in \cite{izotov2011}. {\tt hab1} and {\tt hab11} show much lower equivalent widths, with peak EW(\hb )$\approx 30$ \AA . They can be more normal dwarfs, or tidal dwarfs/knots, as we discussed previously. The distribution of system 11 in BPT diagrams are particularly interesting.
The log([O\,{\sc iii}]/H$\beta$) 
in this system shows relatively small variation.
According to its optical images, 
it is in a merging system with a massive red central galaxy, similar to the galaxies in the second group.
However, as we will see in the next section, the chemical abundance of this system shows a similar pattern to those galaxies in the first group (i.e. low-mass blue galaxies).

The rest of the galaxy systems in our sample (6 and 13) are short of strong emission 
lines other than \ha\ line, thus showing only a few \ha\ blob spaxels in BPT diagrams. Their central galaxies are line-less and they are more similar to systems in the second group with retired host galaxies.

\begin{figure*}
	\includegraphics[width=0.95\textwidth]{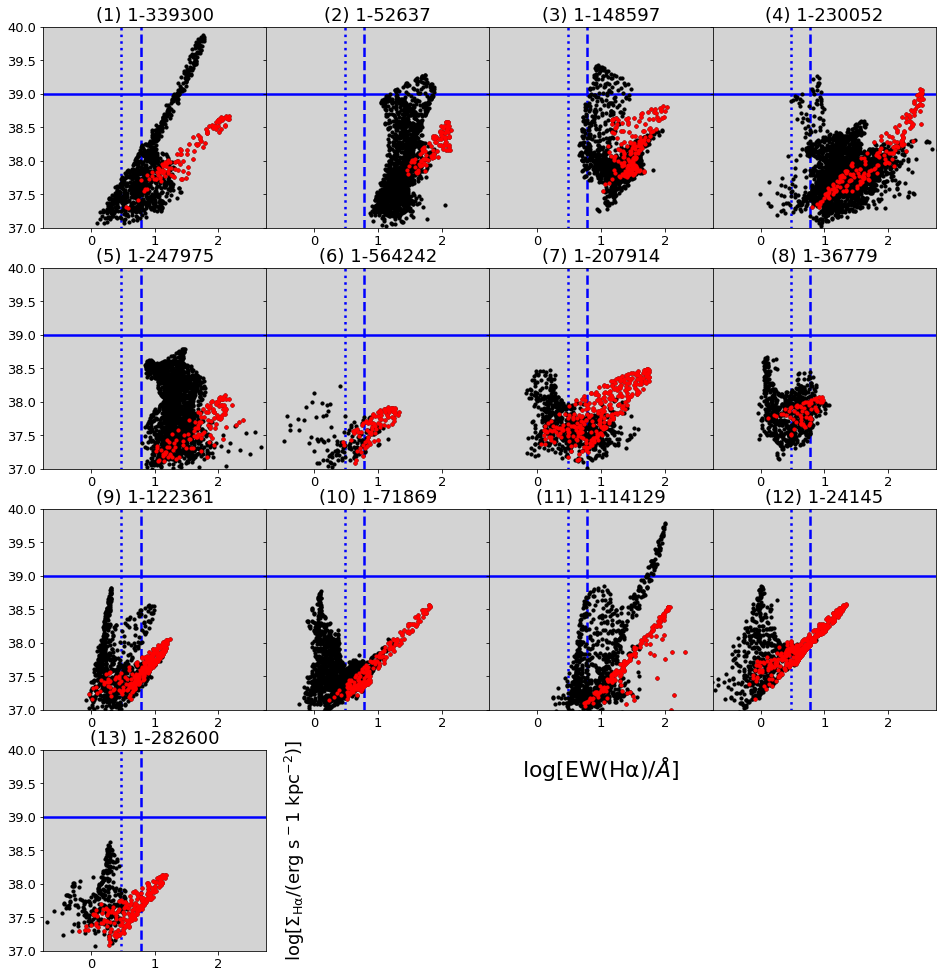}
    \caption{\ha\ surface brightness and \ha\ equivalent width of our sample spaxels. Red and black points correspond to \ha\ blobs and their host galaxies, respectively. The solid blue line corresponds to $\rm \Sigma _{H\alpha} = 10^{39}~erg~s^{-1}~kpc^{-2}$, while the dotted blue line and the dashed blue line correspond to $\rm EW(H\alpha) = 3.0~$\AA\ and $\rm EW(H\alpha) = 6.0~$\AA , respectively.}
    \label{fig:HaF_EW}
\end{figure*}

We note that there is no clear sign of AGN in our sample. Although some galaxy 
spaxels are located in the AGN region in the [N\,{\sc ii}]-based diagram 
(e.g. {\tt gal7}, {\tt gal8}, {\tt gal9}, {\tt gal10}, {\tt gal12}), they are mostly 
classified as LI(N)ERs according to the [S\,{\sc ii}]-based diagram. As shown by
many authors, LI(N)ERs could result from a variety of ionizing sources other than AGN
\citep[e.g.][]{1994A&A...292...13B, 1995ApJ...455..468D, 2010MNRAS.402.2187S, 2011MNRAS.413.1687C,
2012ApJ...747...61Y, 2016MNRAS.461.3111B}.
Shocks could potentially contribute to the ionization in our sample galaxies, 
especially for those merging or undergoing strong SF activities. As we will see in \S~\ref{sec:kinematics}, the low gas velocity dispersions found inside \ha\ blobs disfavor a shock dominated scenario. 
To explore potential contribution from DIG to the total \ha\ fluxes we see, we 
examine the equivalent width as well as the surface brightness of \ha\ line in our 
sample. In \autoref{fig:HaF_EW} we plot the equivalent widths of the \ha\ line, 
EW(H$\alpha$), versus the H$\alpha$ surface brightness, $\rm \Sigma_{H\alpha}$, 
using our sample spaxels. The \ha\ equivalent width measures the line strength 
relative to the underlying continuum, which could be considered as a proxy to the 
specific star formation rate (sSFR) of the region. As expected, we find that \ha\ 
blobs tend to have higher EW(H$\alpha$) than their host galaxies. The surface brightness 
of the \ha\ line, on the other hand, probes the intensity of SF activities, if the 
region is confirmed to be ionized by young stellar populations. We see that \ha\ blobs 
also exhibit comparable or higher $\rm \Sigma_{H\alpha}$ than their host galaxies, 
and quite often we find they form one of the two distinct branches in these figures.

We can use \autoref{fig:HaF_EW} to further understand the ionizing sources, as both 
of its axes are useful in probing contamination from DIG. As shown by \citet{Zhang-17}, 
a lower limit in the \ha\ surface brightness, e.g. $\rm \Sigma_{H\alpha}>10^{39}$erg 
s$^{-1}$kpc$^{-2}$, can be used to isolate H\,{\sc ii} regions (but we note that the actual value for the cut might well depend on the physical conditions of different parts of galaxies, therefore the cut we apply here should not be considered as an exact one). From~\autoref{fig:HaF_EW}, 
we see that the $\rm \Sigma_{H\alpha}$ of most our sample galaxies and \ha\ blobs 
does not meet this criterion, suggesting that they either have contamination from DIG, 
or their SFR is considerably low. Unlike in \autoref{tab:global}, 
we do not apply the extinction correction to the data points here, as the Balmer 
decrement method could potentially overestimate the extinction in regions having 
DIG \citep{2018MNRAS.481..476Y}. Even after we include the extinction correction, 
nearly all of the \ha\ spaxels still lie below this demarcation. EW(H$\alpha$) is 
also useful in selecting pure SF regions. We take two demarcation lines with 
EW(H$\alpha$) = 3.0 \AA\ and EW(H$\alpha$) = 6.0 \AA\ from \cite{2011MNRAS.413.1687C}, 
which were originally combined with log([N\,{\sc ii}/H$\alpha$]) to form so-called 
WHAN diagram. Active SF or AGN usually produce EW(H$\alpha$) higher than 6.0 \AA,
whereas retired galaxies tend to show EW(H$\alpha$) well below 3.0 \AA. We can see 
that the AGN spaxels identified in {\tt gal7}, {\tt gal8}, {\tt gal9}, {\tt gal10}, 
and {\tt gal12} actually have low EW(H$\alpha$). As a result, they are probably not 
true AGNs, but DIG. This indicates that galaxies in the second 
group we identified in BPT diagrams are likely inactive or retired. In contrast, 
their \ha\ blobs show EW(H$\alpha$) higher than 6.0 \AA. Therefore, it is possible 
that these \ha\ blobs are still forming stars, although at a relatively low level. 
The first group of galaxy systems we identified in BPT diagrams (1, 2, 3, 4, and 5) 
show signatures of active SF, with the \ha\ blobs having the highest sSFR. The 
merging system 11 also shows similar level of SF. Overall, despite the relatively 
low $\rm \Sigma_{H\alpha}$, there appear to be an enhancement of SF activities in \ha\ blobs.

We could estimate the upper limit of the SFR in our sample using $\rm \Sigma_{H\alpha}$.
Applying the conversion factor given by \cite{1998ARA&A..36..189K}, we find that the SFR surface densities of the \ha\ blobs are mostly lower 
than $\sim$ 10 $^{-3} \text{M}_\odot \ \text{kpc}^{-2} \ \text{yr}^{-1}$,
and the upper limits of the integrated SFR to be lower than 
$\sim$ 0.2 M$_\odot$ yr$^{-1}$, with a median value of $\sim$ 0.03 M$_\odot$ yr$^{-1}$. 
Given the low level of star formation and the lack of AGN in our sample, we argue that 
the \ha\ blobs are unlikely to be produced by AGN or star formation driven outflows.

\subsection{Gas-phase metallicities}
\label{sec:MZR}

%\begin{figure*}
%	\includegraphics[width=0.95\textwidth]{Section3/New_Fig4.png}
%    \caption{Spatially resolved gas-phase metallicity maps of H\,{\sc ii} regions in the sample galaxies. The locations of \ha\ blobs are indicated by black contours. The solid and dashed contours correspond to the $\rm 95^{th}$ and $\rm 63^{rd}$ percentiles of the \ha\ surface brightness in \ha\ blobs.}
%    \label{fig:resolved_z}
%\end{figure*}

\begin{figure*}
	\includegraphics[width=0.9\textwidth]{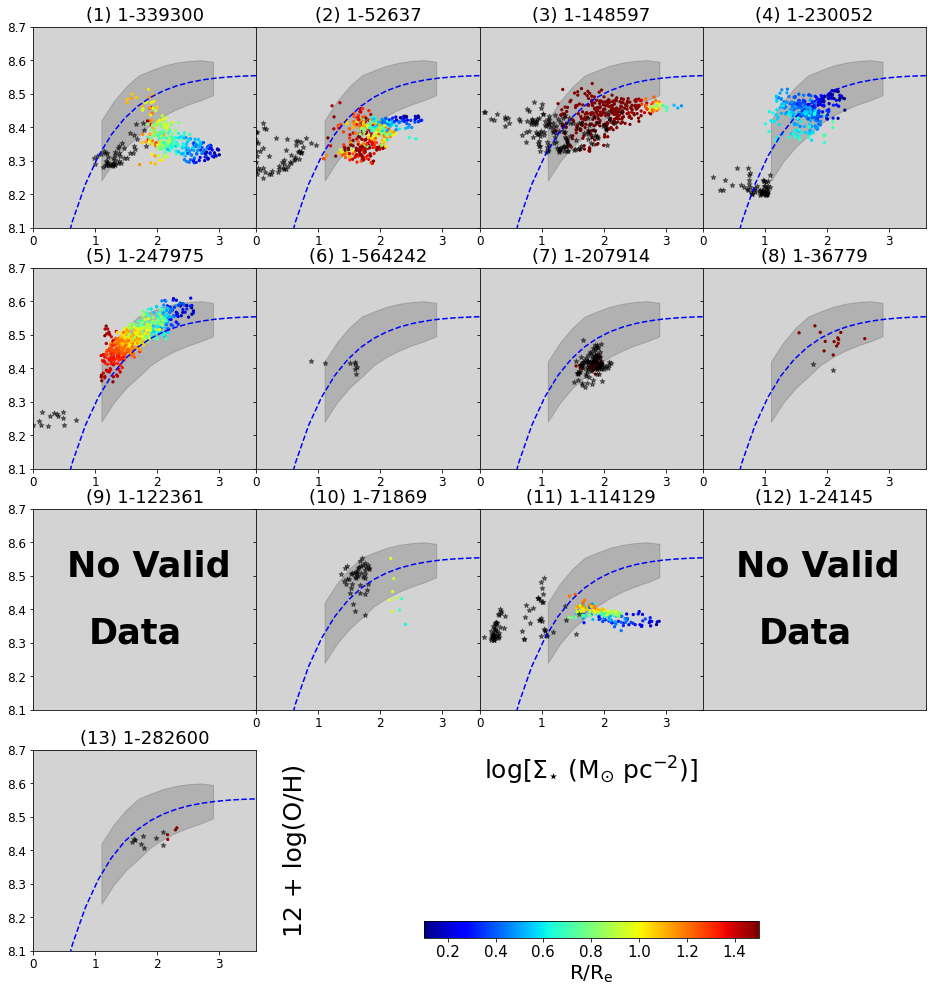}
    \caption{Spatially resolved MZR of our sample spaxels. The metallicity is obtained through the O3N2 method. The stellar masses of the host galaxies are computed by the spectral fitting code {\tt pPXF}, and the stellar masses of the \ha\ blobs are estimated through the SED fitting code {\tt CIGALE} using the MaNGA synthesized $g$, $r$, and $z$ band fluxes. While the host galaxy spaxels are color-coded according to their radial positions, R/R$\rm _{e}$, the \ha\ blob spaxels are colored in black. 
    The blue dashed line represents the median trend with extrapolations at low and high surface mass density ends.
    The shaded region on the background describes the $1\sigma$ variation in metallicity.}
    \label{fig:a2}
\end{figure*}

\begin{figure*}
	\includegraphics[width=0.95\textwidth]{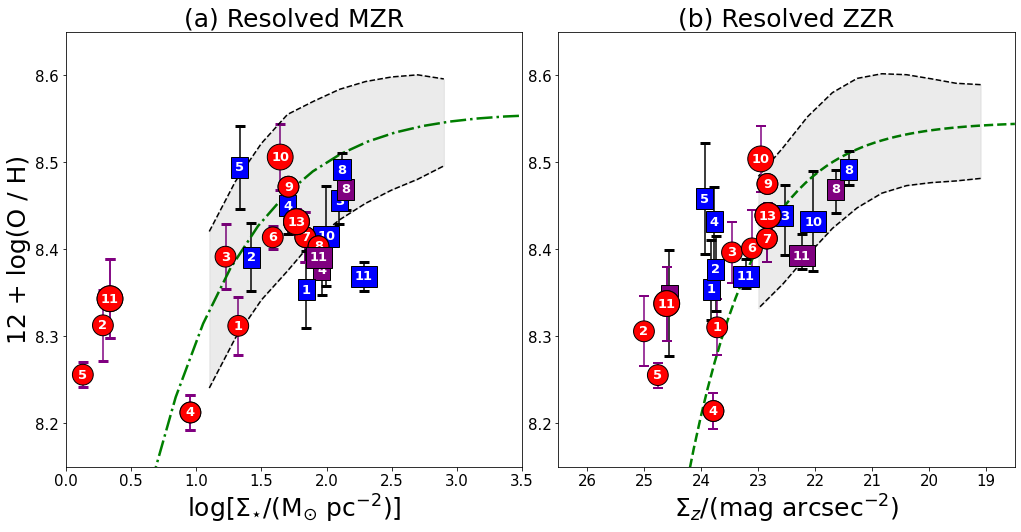}
    \caption{Left panel: resolved mass-metallicity relation of \ha\ blobs and their host galaxies. For each galaxy and \ha\ blob, we only plot their median values.
    The red circles are \ha\ blobs and the blue boxes are their host galaxies. Purple boxes are other galaxy members observed inside the MaNGA IFU bundles. Galaxies/\ha\ blobs that belong to the same system are marked with the same serial number. Error bars indicate the $1 \sigma$ variations of the metallicity inside the blobs/galaxies. The green dash-dotted line shows the fitted median trend with extrapolations at low and high surface mass density ends. The two sets of black dashed lines enclose spaxels within one standard deviation from the median along the vertical axis. Right panel: resolved metallicity - $z$-band surface brightness relation of \ha\ blobs and their host galaxies.
    }
    \label{fig:Global_MZR_ZD}
\end{figure*}

The gas-phase metallicity is closely related to the SFH of a galaxy, 
as it encodes the accumulated chemical abundance over the entire 
evolution history of the galaxy. The scaling relation between gas-phase 
metallicities and stellar masses of star-forming (SF) galaxies, 
known as the mass-metallicity relation (MZR), has long been studied
and provides a general reference for the chemical evolution history 
of SF galaxies or SF regions \citep{2004ApJ...613..898T}. In this 
section, we examine the gas-phase metallicities of our sample galaxies 
and \ha\ blobs from a spatially resolved perspective.

%\autoref{fig:resolved_z}
\autoref{fig:a2}
shows the spatially resolved MZR for the 13 galaxy systems.
We estimate the stellar mass of each spaxel within the host galaxies by applying the spectral fitting code {\tt pPXF} \citep{2004PASP..116..138C, 2017MNRAS.466..798C} to the MaNGA spectra. 
The stellar templates we use are drawn from the MILES stellar-template library \citep{sanchez-blazuez2006,falcon-barroso2011}.
For \ha\ blobs, however, this method is not applicable due to the low S/N of the spectra. We opt for {\tt CIGALE} \citep{2005MNRAS.360.1413B} and estimate stellar masses for spaxels inside the blobs by fitting the spectral energy distributions (SEDs) using fluxes in $g$, $r$, and $z$ bands. The fluxes are drawn from the MaNGA DRP products. A consistency check has been performed to ensure that {\tt pPXF} and {\tt CIGALE} provide identical stellar mass estimates for the host galaxies of the \ha\ blobs. When calculating the surface mass densities for the host galaxies of \ha\ blobs, we correct for the effect of inclination according to the minor-to-major axis ratios measured in $r$-band images by the SDSS photometric pipeline.

The gas-phase metallicities (defined as 12 + log(O/H)) of our 
sample galaxies are calculated with the O3N2 method 
\citep{2013A&A...559A.114M}, which is based on measurements of strong 
optical emission lines. Below is the equation given by \cite{2013A&A...559A.114M},
\begin{equation}
    \rm 12+log(O/H) = 8.533 - 0.214\cdot \log ( \frac{[O~{\sc III}]\lambda5007/H\beta } 
    {[N~{\sc II}]\lambda6584/H\alpha}).
\end{equation}
This method is calibrated with 
observations of H\,{\sc ii} regions and is sensitive to the ionization 
parameter of the ionized cloud. Therefore, we only consider spaxels that 
are identified as H\,{\sc ii} regions in the optical diagnostic diagrams 
(\citealp{1981PASP...93....5B, 1987ApJS...63..295V, 2003MNRAS.346.1055K, 
2006MNRAS.372..961K}). Most of the \ha\ blobs in our sample are classified 
as H\,{\sc ii} regions, according to our analyses in \S~\ref{sec:photo}. In addition, 
we require the relevant emission lines, i.e. [O {\sc iii}]$\lambda$5007, 
H$\beta$, [N {\sc ii}]$\lambda$6583, and H$\alpha$, to have signal-to-noise 
ratio (S/N)$>3$.

We note that there are usually significant and 
unavoidable systematic differences between the results from different 
metallicity estimators \citep[see section 4 of][]{2019ARA&A..57..511K}.
The reasons we choose the O3N2 method are twofold. On the one hand, this method only uses strong emission lines and is insensitive to the dust extinction. On the other hand, this method was also adopted by \cite{2016MNRAS.463.2513B} to derive the spatially resolved MZR with MaNGA data, thus facilitating the comparison we make here.
\cite{2016MNRAS.463.2513B} derived a spatially resolved MZR using 653 disk galaxies in MaNGA. The dotted-dashed line in \autoref{fig:a2} represents the median trend of the spatially resolved MZR, and the shaded region shows the $1 \sigma$ variation in metallicity. This relation is truncated at both low mass and high mass end, and we have extrapolated the median trend as some of our sample spaxels fall into these regimes.
In what follows, we will focus on the relative differences of 
the metallicity in our sample, rather than their absolute values.

One can see from the \autoref{fig:a2} that we have obtained metalliciy 
measurements for most of the \ha\ blobs, indicating significant detections 
of all four emission lines. In the case of {\tt hab9} and {\tt hab12}, however, we are not able to get reliable measurements
of their metallicities as most of their spaxels are found in the composite 
region in BPT diagrams. 
In most cases where we have measured gas-phase metallicities for both 
the \ha\ blob and the host galaxy, the \ha\ blob shows relatively low 
metallicities compared to their host galaxies 
(e.g., galaxy system No. 2, 3, 4, 5, 8). In galaxy system 1 and 11, 
the derived metallicity shows little variation across the whole system 
(except near the edge of the galaxy), and the metallicity of the \ha\ blob 
is similar to that of the galactic center. The $\rm 10^{th}$ system is the 
only case where the \ha\ blob appears much more metal-rich than the host galaxy, 
but we cannot exclude the possibility that there is contamination from the DIG, and there are very few spaxels in the host galaxy that are identified as H\,{\sc ii} regions. H\,{\sc ii} regions in system 6, 7, and 13 are also mostly found in their \ha\ blobs.

We notice that the galaxies and the \ha\ blobs as a whole cover a relatively narrow range of metallicity, limited to intermediate-to-low values. In contrast, their surface mass densities and flux densities span several orders of magnitude, indicating a flattening of radial metallicity profiles of our sample galaxies, as compared to other galaxies. In \autoref{fig:a2}, we color code the galaxy spaxels according to their relative radial distances to the center of the galaxies (scaled by the effective radii measured in $r$-band), while plotting the spaxels at the locations of the \ha\ blobs in black. We can see that only a few galaxies follow the average relation of the general population ({\tt gal4} and {\tt gal5}), and most galaxies deviate from the average relation in the sense that their inner regions present high densities but lower-than-average metallicities. As a result, the spaxels at smaller radii fall below the average relation to varying degrees, while the spaxels in the outskirt have similar metallicities, and in some cases even higher metallicities, than the inner region, yielding flat or negative gradients.
The seemingly pre-enriched metallicity of \ha\ blobs is consistent with the picture where they were originally a part of the nearest galaxies. An alternative explanation is the inflow of low-metallicity gas diluting the metallicity in the galaxy, which could also be a result of the interaction between galaxies.

In the left panel of \autoref{fig:Global_MZR_ZD}, we plot the median values of the surface mass density and metallicity of our sample galaxies and \ha\ blobs. For systems with a pair or trio of galaxies, we divide the spaxels into different galaxy components according to their $r$-band images, and color the companion galaxies in purple.
Similar to what we have seen in \autoref{fig:a2}, most \ha\ blobs show lower median metallicities compared to their host galaxies. A few \ha\ blobs with the lowest stellar mass surface densities lie well above the extrapolated median trend of \cite{2016MNRAS.463.2513B}, appearing to be much more metal enriched compared with the general population of SF regions in MaNGA.
But we note that one should be cautious about the interpretations based on the extrapolation of the median trend, due to the potential flattening of the resolved MZR for low mass galaxies \citep[see figure 7 of][]{2016MNRAS.463.2513B}.
The galaxies, on the other hand, tend to systematically fall below the median relation at high stellar mass surface densities.

Considering that stellar mass estimates of the \ha\ blobs may have suffered 
from large uncertainties due to their low spectral S/N, we consider the fluxes in $z$-band as a substitute for the stellar masses. 
In the right panel of \autoref{fig:Global_MZR_ZD}, we show the relation 
between gas-phase metallicity and the $z$-band surface brightness, $\Sigma_z$, 
for both the galaxies and \ha\ blobs. 
The $z$-band surface brightness can serve as a good proxy of the stellar mass, 
as it is characterized by star lights from longer wavelength in the infrared 
compared with other SDSS photometric bands and exhibits relatively small 
variations in the mass-to-light ratios \citep{2001ApJ...550..212B, 2003MNRAS.341...33K}.
Following the work by \cite{2016MNRAS.463.2513B}, we derive a 
`$z$-band surface brightness vs. metallicity relation', or ZZR, using 
H\,{\sc ii} regions in MPL-7. These H\,{\sc ii} regions are identified 
using the [N\,{\sc ii}] BPT diagram, and we obtain a total of $\sim 1.8\times 10^6$ spaxels.
Overall, the results are similar to what are shown in the left panel, 
despite that the galaxies seem slightly closer to the 
median trend of the general population.

Since our \ha\ blobs are all found within the MaNGA's hexagonal field of views and lie close to their host galaxies, there is potential contamination from the star lights of the stellar halos of the central galaxies, which makes us overestimate the stellar masses of \ha\ blobs. To investigate this effect, we use the model-subtracted optical images provided by the Legacy Surveys (see Appendix~\ref{model_sub}). We recompute the stellar masses with the residual fluxes at the locations of \ha\ blobs after the stellar components of the central galaxies are subtracted. This approach in general lowers the estimated stellar mass surface densities of \ha\ blobs by $0.5\sim 1$ dex (and $1\sim 2$ magnitude for the $z-$band surface brightness). If we assume that the star lights associated with \ha\ blobs are all within the residual images, then most of the \ha\ blobs would be highly chemically enriched in comparison with the typical SF regions. However, it is unclear whether this assumption holds, especially for those \ha\ blobs in the first group of the galaxies that appear as off-galaxy SF regions. Moreover, the derivation of the stellar masses based on residual fluxes rely heavily on the accuracy of the source subtraction. In general, we can consider the stellar masses for \ha\ blobs in \autoref{fig:a2} as the upper limits. Whereas their true stellar masses can be similar to or even lower than the values derived from residual fluxes.

If \ha\ blobs are normal SF regions or dwarf galaxies, we expect their MZRs to be compatible with the general trend. While for \ha\ blobs associated with tidal remnants, despite that they are likely chemically pre-enriched (as they originate from their nearby galaxies), whether the observed MZRs will deviate from the general trend also depends on how massive the associated stellar component is, as well as how the stellar mass surface density changes as the tidal feature evolves. Moreover, gas inflows during mergers might influence the elemental abundances of these \ha\ blobs as well, which is indicated by the seemingly flattened metallicity gradients in some of the sample galaxies. Therefore, one should be cautious while interpreting the MZRs found in these systems. Regardless, the bottom line conclusion is that if the MZR of the \ha\ blob significantly lies above the general trend, then it is likely related to tidal remnants. On the other hand, if the MZR of the \ha\ blob fits into the general trend, both scenarios (normal SF regions or tidal remnants) is plausible.

\subsection{Kinematics of ionized gas}
\label{sec:kinematics}

\begin{figure*}
	\includegraphics[width=0.95\textwidth]{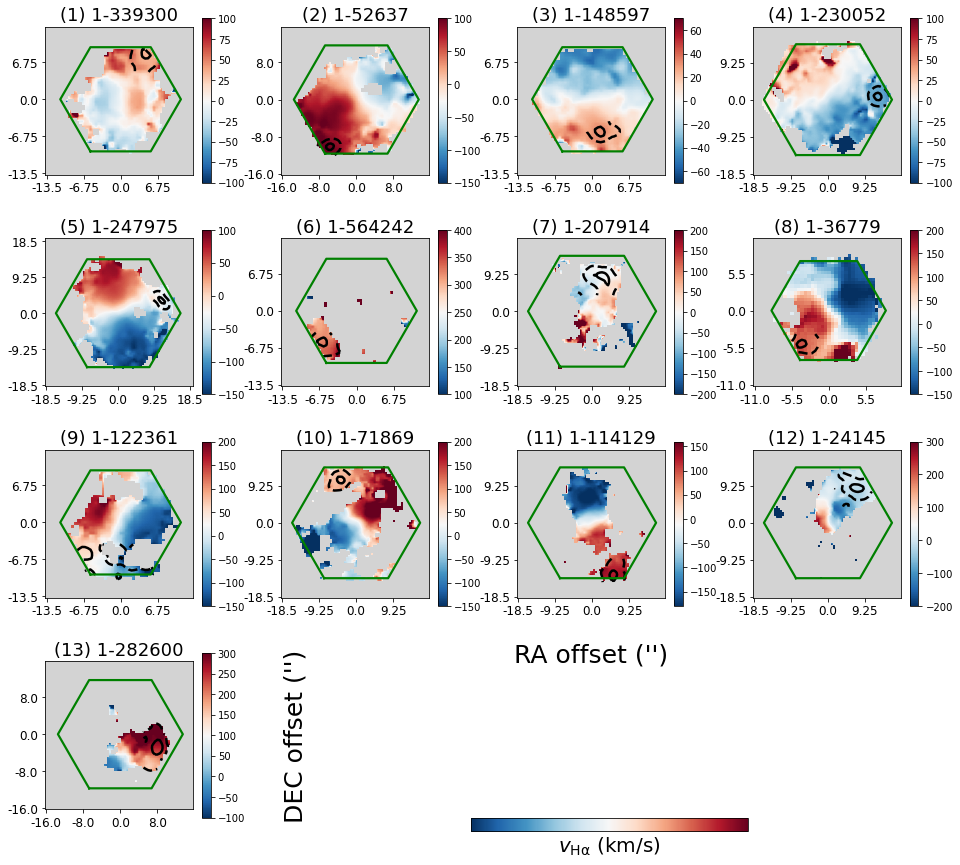}
    \caption{Velocity fields traced by the \ha\ line in the sample galaxy systems. The locations of \ha\ blobs are indicated by black contours. The solid and dashed contours correspond to the $\rm 95^{th}$ and $\rm 63^{rd}$ percentiles of the \ha\ surface brightness in \ha\ blobs.}
    \label{fig:resolved_vel}
\end{figure*}

\begin{figure*}
	\includegraphics[width=0.95\textwidth]{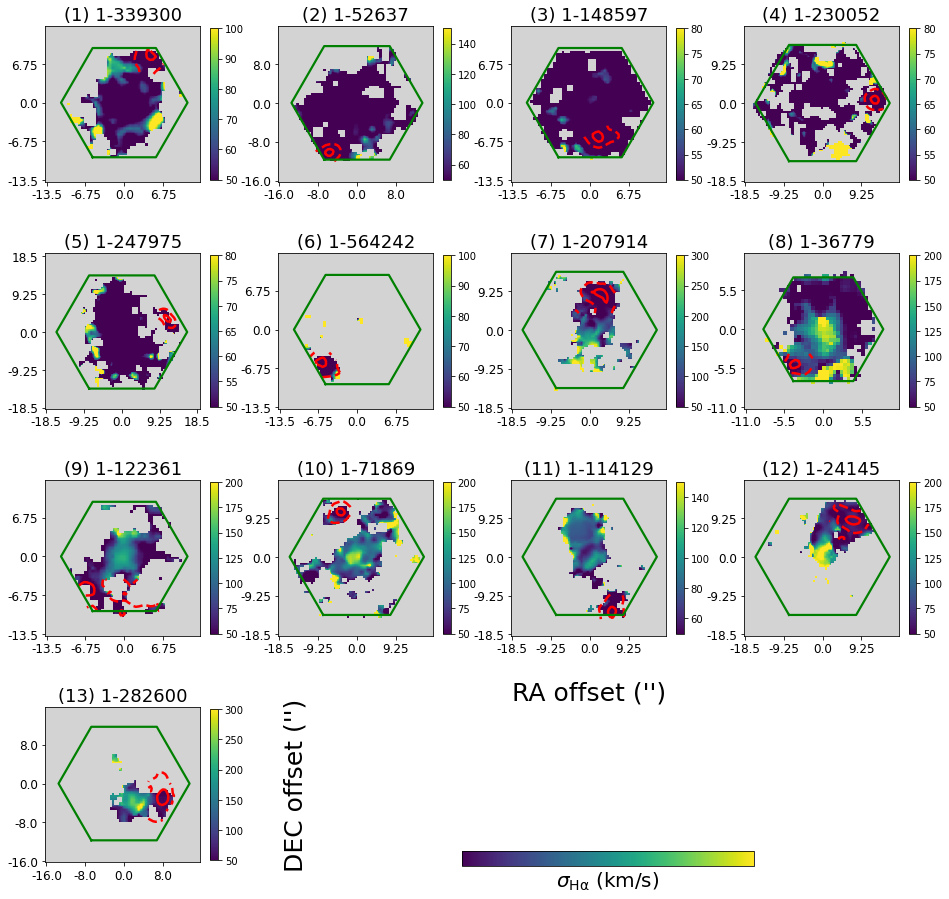}
    \caption{Velocity dispersion maps traced by the \ha\ line in the sample galaxy systems. The locations of \ha\ blobs are indicated by red contours. The solid and dashed contours correspond to the $\rm 95^{th}$ and $\rm 63^{rd}$ percentiles of the \ha\ surface brightness in \ha\ blobs.}
    \label{fig:resolved_sig}
\end{figure*}

The kinematics of the ionized gas could provide information about the current relation between the \ha\ blobs and their nearest galaxies.

\begin{figure}
	\includegraphics[width=0.47\textwidth]{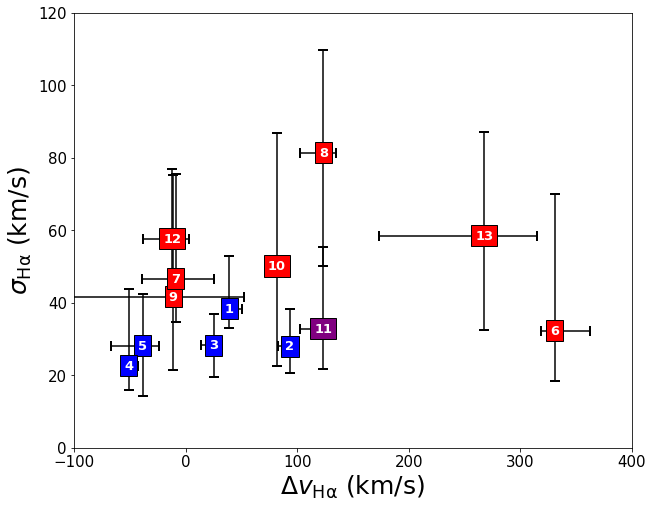}
    \caption{\ha\ velocity versus \ha\ velocity dispersion for \ha\ blobs in our sample. The colored squares with number show the median values among the spaxels within \ha\ blobs, and the error bars enclose the middle 84\% of the spaxels. While the first group of \ha\ blobs are colored in blue, the second group of \ha\ blobs are colored in red. {\tt hab11} is colored in purple, as system 11 shares certain similarities with both groups.}
    \label{fig:kinematics}
\end{figure}

\autoref{fig:resolved_vel} shows the velocity fields
of the ionized gas in our sample traced by the \ha\ line, and \autoref{fig:resolved_sig} 
shows the intrinsic velocity dispersion of the \ha\ line (with the instrumental velocity dispersion subtracted). Overall, the \ha\ blobs 
have very similar velocities to the nearby regions of the galaxies, despite 
that they appear physically separate in optical images. Purely judging 
from the velocity maps, the \ha\ blobs do not seem to be isolated components in galaxy systems. \ha\ blobs also exhibit low velocity dispersions 
typical of H\,{\sc ii} regions, with $\sigma _{\rm H\alpha} \lesssim \rm 50~km/s$ 
\citep{2010MNRAS.401.2113E}. In comparison, shocks produced by galactic winds 
or mergers are often featured by much higher velocity dispersions, with 
$\sigma _{\rm H\alpha} \sim \rm 150-500~km/s$ \citep[see][and references 
therein]{2019ARA&A..57..511K}. Interestingly, the gas velocity dispersions 
of systems consisting of multiple galaxies appear to be higher, indicative 
of the presence of low velocity shocks or AGN ionization. We can see that 
kinematically speaking, galaxy system 6, 7, 8, 9, 10, 11, 12, and 13 share 
the similarity that their main galaxies show higher velocity dispersion, 
and many of them have spatially close companions. On the other 
hand, galaxy system 1, 2, 3, 4, and 5 exhibit low velocity dispersions 
across the entire system, which is consistent with what we see in \autoref{fig:mass_color}. 
Therefore, kinematically speaking, these low-mass blue galaxies (or group 1, which we introduced back in \S~\ref{sec:Selection}) seem to form a distinct group as well. Despite the
differences in the host galaxies, all \ha\ blobs in our sample show 
low velocity dispersions, and low to median line-of-sight velocities relative 
to their central galaxies, as shown by \autoref{fig:kinematics}. Although the \ha\ blobs in group 2 exhibit slightly larger velocity dispersions, only {\tt hab8} reaches 80 km/s. {\tt hab6} and {\tt hab13} are the only two have relatively large velocities close to $\sim 300$ km/s, but their median velocity dispersions are still below 60 km/s. This largely 
rules out the ram-pressure stripping scenario as the diffuse emission from 
the stripped gas usually exhibits both high velocity dispersion ($\sigma _{\rm H\alpha} > \rm 100~km/s$) 
and high velocities \citep[e.g.][although the velocity dispersion of the 
dense knots in the ram-pressure-stripped gas could be lower than 50 km/s]{2017ApJ...844...49B}.

The presence of shock ionization may be identified by separate (and possibly broad) 
kinematic components in emission lines. Unfortunately, MaNGA's spectral 
resolution ($\rm \sim 70~km/s$) is not high enough to separate the potential 
shock component. But one could check the correlation between the shock-sensitive 
line ratios and the measured velocity dispersions, which is another evidence 
for shock ionization \citep[e.g.][]{2011ApJ...734...87R, 2014MNRAS.444.3894H, 
2015ApJS..221...28R}. We examine the correlation between [N\,{\sc ii}]/H$\alpha$ 
(as [N\,{\sc ii}]$\lambda 6583$ are detected in all of our sample galaxies and 
show relatively high S/N) and velocity dispersion of \ha\ line. We find that 
there are positive correlations in most galaxies, especially for those having 
higher $\sigma _{\rm H\alpha}$. However, due to the low spatial resolution of MaNGA, 
it is possible that this correlation is produced by the spatial mixing of 
H\,{\sc ii} regions and AGN regions (or DIG). We indeed see that in these systems, 
the $\sigma _{\rm H\alpha}$ is also anti-correlated with the radial position, 
implying potential correlation with AGN ionization or evolved stellar population. 
Since the AGN ionized regions and DIG that resides in pressure-supported regions can also 
have $\sigma _{\rm H\alpha}$ higher than 100 km/s, the contamination from them 
could contribute to the correlation we observed in MaNGA. In summary, although 
it is plausible that shocks play an important role in systems hosting \ha\ blobs, 
we need higher spectral resolution as well as spatial resolution to verify this scenario.

\section{Discussion}
\label{sec:discussion}

In this section, we compare our work with previous studies of \ha\ blobs, and 
we attempt to propose a consistent picture for the origin of \ha\ blobs in our 
sample based on the results presented in the previous section. The derived properties of \ha\ blobs in this work are summarized in \autoref{tab:derived}.

\subsection{The physical origin of \ha\ blobs}

\begin{table*}
%	\centering
	\caption{Derived properties of \ha\ blobs}
	\label{tab:derived}
	\begin{tabular}{lccccccccc} 
		\hline
		\hline
		No. & Morphology\mbox{*} & No. of companion  & BPT &  12 + log(O/H) & $\log(\Sigma _*)$\mbox{**} & $\Sigma _z$\mbox{**} & $\rm \Delta v_{H\alpha}$ & $\rm \sigma _{H\alpha}$ \\
                    &          &     galaxies      & classification   & (via O3N2) & ($\rm \log(M_{\odot}~pc^{-2})$)  & ($\rm mag~arcsec^{-2}$) & (km/s) & (km/s) \\
		\hline
		1 & spot & 1 & H\,{\sc ii} & $8.31\pm0.03$ & $1.32(0.80)\pm0.88$ & $23.7(25.4)\pm0.9$ & $39\pm12$ & $38\pm12$ \\
		2 & tail/spiral & 1 & H\,{\sc ii} & $8.31\pm0.04$ & 0.28(-)$\pm0.34$ & 25.0(-)$\pm0.3$ & $94\pm8$ & $28\pm8$ \\
		3 & faint disk & 1 & H\,{\sc ii} & $8.40\pm0.03$ & 1.22(-)$\pm0.45$ & 23.5(-)$\pm0.9$ & $25\pm8$ & $28\pm8$ \\
		4 & spot & 1 & H\,{\sc ii} & $8.21\pm0.02$ & $0.96(0.96)\pm0.15$ & $23.8(23.8)\pm0.2$ & $-51\pm12$ & $23\pm22$ \\
		5 & tail/spiral & 1 & H\,{\sc ii} & $8.26\pm0.01$ & 0.13(-)$\pm0.30$ & 24.8(-)$\pm0.4$ & $-38\pm23$ & $28\pm33$ \\
		6 & - & 2 & H\,{\sc ii} & $8.40\pm0.04$ & $1.59(0.74)\pm0.74$ & $23.1(25.7)\pm1.0$ & $331\pm23$ & $32\pm31$ \\
		7 & tail/spiral & 1 & H\,{\sc ii} & $8.41\pm0.03$ & $1.84(0.66)\pm0.80$ & $22.8(24.9)\pm1.0$ & $-9\pm32$ & $47\pm29$ \\
		8 & ring & 2 & composite & - & $1.94(0.48)\pm0.63$  & $22.1(26.6)\pm1.2$ & $124\pm18$ & $81\pm36$ \\
		9 & ring & 1 & composite & - & 1.92(-)$\pm0.05$ & 22.8(-)$\pm0.3$ & $-11\pm69$ & $42\pm34$ \\
		10 & faint disk  & 1 & H\,{\sc ii} & $8.50\pm0.04$ & $1.64(1.13)\pm0.77$ & $23.0(24.4)\pm0.8$ & $82\pm18$ & $50\pm33$ \\
		11 & spot & 2 & H\,{\sc ii} & $8.33\pm0.04$ & $0.34(0.34)\pm0.40$ & $24.6(24.6)\pm0.6$ & $123\pm17$ & $33\pm20$ \\
		12 & - & 2 & composite & - & 2.41(-)$\pm0.06$ & 21.7(-)$\pm0.3$ & $-12\pm30$ & $58\pm24$ \\
		13 & - & 3 & H\,{\sc ii} & $8.44\pm0.01$ & $1.77(1.21)\pm0.79$ & $22.8(23.8)\pm0.7$ & $268\pm60$ & $58\pm33$ \\
		\hline
	\end{tabular}
	\begin{tablenotes}
%	    \centering
        \small
        \item \mbox{*} This is the morphology of the underlying optical structures at the locations of the \ha\ blobs revealed by the deep images, which are not necessarily the optical counterparts of these \ha\ blobs.
        \item \mbox{**} These values should be considered as upper limits. Values shown in parentheses are model-subtracted values (see Appendix~\ref{model_sub}).
    \end{tablenotes}
\end{table*}

For the first discovered \ha\ blob, i.e. {\tt hab12} in our sample, 
\cite{Lin-17} concluded that their analyses showed that there are 
two equally possible scenarios: (1) it is an unusual dwarf galaxy or a gas-rich ultra-diffuse galaxy 
(UDG) that is interacting with the central galaxy; (2) it is a gas cloud ejected 
by the past AGN activity of the central galaxy, and is possibly still being ionized 
by a weak AGN. The most recent follow-up of this \ha\ blob by \cite{Pan-20} with X-ray observations, 
however, led to a different conclusion that this \ha\ blob is probably formed 
by the cooling of the hot inter-galactic medium (IGM), consistent with the 
findings of \cite{2019MNRAS.488.2925O}. This was based on the following two 
results found in \cite{Pan-20}: (1) the blob exhibits scaling relations different 
from what are observed in the nearby galaxies, and (2) there is no sign of ongoing 
AGN activity. In this work, since we only use data in the optical, we are unable 
to make a solid conclusion about {\tt hab12} like in \citet{Pan-20}. However, we 
have seen that about half of the \ha\ blobs in our sample show a number of similarities 
to {\tt hab12}. The galaxy system 6, 7, 8, 9, 10, 12, and 13 all show signatures of 
interactions, similar kinematics, and similar ionizing patterns. The galaxy system 
11, on the other hand, appears as the gas-rich counterparts of the above systems. 
It is plausible that all the \ha\ blobs in these galaxy systems share the same origin. 
But we note that there is significant difference in the large scale environment 
between galaxy system 12 and other galaxy systems in our sample, as the former 
is involved in a group-group merger and is close to the brightest galaxy of the 
other group \citep{2019MNRAS.488.2925O}. Most of the \ha\ blobs in our sample 
are not found in such a dense environment. In addition, the ionizing SED of the 
{\tt hab12} seems harder compared with most of the other \ha\ blobs, as its spaxels 
are identified as composites rather than H\,{\sc ii} regions in the [N\,{\sc ii}] 
BPT diagram. Although {\tt hab6}, {\tt hab7}, {\tt hab8}, {\tt hab9}, and {\tt hab13}
also have composite spaxels, the number of such regions is considerably smaller. 
Therefore, it is unlikely that the cooling IGM is the true identity for the 
majority of the \ha\ blobs in the second group.

Alternative explanations for these \ha\ blobs include disturbed gas-rich UDGs, AGN ejected 
clouds, and tidal remnants. Since most \ha\ blobs in our sample appear kinematically 
connected to their central galaxies and show similar rotation velocities to the nearby 
galaxy regions, they are unlikely to be independent galaxies (although this could be a result of bias in sample selection, which we will discuss in \S~\ref{bias}). The work by \cite{Bait-Wadadekar-Barway-19} supported both the tidal stripping and the AGN scenarios, which reported on the discovery of 6 
\ha\ blobs in the MaNGA galaxies using the public data of SDSS DR14. Owing to the 
different sample selection method, their sample have a small overlap with ours, 
with only one blob ({\tt hab6}) that is included in both samples. We will discuss the sample selection difference and our interpretation of Bait et al.'s sample in Appendix~\ref{bait}. Since all 
of the galaxies in the second group of our sample exhibit \ha\ equivalent widths similar 
to retired regions or diffuse ionized gas (DIG), the existence of ongoing 
AGN activities seems implausible. After excluding these two possibilities, it seems 
that the tidal interaction scenario is the most likely one. \ha\ blobs in the second 
group could originally be part of the central galaxies, but stripped out in the past 
due to the tidal interaction. Their seemingly pre-enriched chemical abundance is another evidence 
for this scenario, similar to what have been observed in tidal dwarfs 
\citep{1998A&A...333..813D, 2003A&A...397..545W}.

In contrast, the first group of galaxies and \ha\ blobs in our sample (galaxy system 
1, 2, 3, 4, and 5) have very different properties. These systems are featured by their 
low masses, blue colors, and ongoing star formation activity across the entire system. 
In addition, they have no obvious close companions or obvious signs of interactions. 
While the \ha\ blobs in these systems appear to follow the mass-metallicity relation 
(MZR) of nearby galaxies, their host galaxies show a flattened metallicity gradient. 
This could be due to accretion or inflow of low-metallicity gas. Combining these facts, 
we argue that these \ha\ blobs are likely clumps of star-forming regions or gas-rich 
dwarf galaxies that are visually offset from the central galaxies in MaNGA IFU. If the 
former is true, they might be associated with very faint disks or spirals, which is broadly 
consistent with the morphological identification we made in \autoref{fig:deep_imaging_e}.

\begin{figure}
	\includegraphics[width=0.47\textwidth]{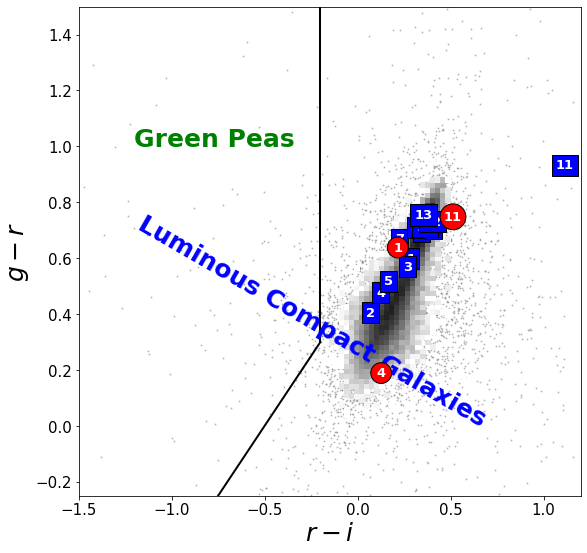}
    \caption{$g-r$ vs. $r-i$ diagram. The host galaxies of \ha\ blobs are plotted as blue rectangles and \ha\ blobs with spot-like optical counterparts are plotted as red circles. A representative sample of nearby galaxies drawn from NSA catalogue are shown in the background. The demarcation lines defined by \protect\cite{cardamone2009} are used to separate green pea galaxies from normal galaxies. The sequence of luminous compact galaxies found by \protect\cite{izotov2011} is indicated by the text.}
    \label{fig:cc_diagram}
\end{figure}

Specifically, we would like to comment on \ha\ blobs with clearly detected spot-like optical counterparts, i.e. {\tt hab1}, {\tt hab4}, and also {\tt hab11}. The shape of the optical counterpart of {\tt hab4} appear to be most spherical, while those of {\tt hab4} and {\tt hab11} are more irregular and more diffuse. As we have noted, {\tt hab11} lives in a merging system, and is more chemically-enriched compared to average SF regions. On the other hand, spaxels in {\tt hab1} and {\tt hab4} show MZR consistent with the extrapolated median trend of SF regions (see \autoref{fig:a2}). Therefore, despite their similar optical counterparts, {\tt hab11} is more likely to have a tidal origin. {\tt hab1} and {\tt hab4} both show blue colors and very strong emission lines. Back in \S~\ref{sec:photo}, we have checked their equivalent widths of \hb\ and [O\,{\sc iii}]$\lambda 5007$ lines and rule out that they are extreme emission-line dwarfs like green pea galaxies or their low-mass counterparts, blueberry galaxies \citep{yang2017}. \autoref{fig:cc_diagram} shows the $g-r$ vs. $r-i$ diagram for our sample galaxies and \ha\ blobs. One can see none of these \ha\ blobs fall into the region of green pea galaxies defined by \cite{cardamone2009}. Among the \ha\ blobs, {\tt hab4} has the bluest color and lies in the sequence of luminous compact galaxies (LCGs) found by \cite{izotov2011}. It has the highest equivalent width, with EW(\hb ) > 50 \AA\ . However, its \hb\ luminosity, L(\hb ) < $10^{40}$ erg/s, which does not meet the criterion of LCGs. Therefore, it is likely a faint blue compact dwarf. In comparison, {\tt hab1} appears even fainter and extended. Due to its close distance to its host galaxy, it could be a dwarf galaxy undergoing galaxy interaction. The inverted metallicity gradient in {\tt gal1}, similar to the case of {\tt gal11}, (see \autoref{fig:a2}) might be a result of gas inflow during the interaction.

Finally, one may wonder whether the \ha\ blobs detected in the nearby universe are somehow related to the Ly$\alpha$ blobs at high redshifts \citep[see e.g.][]{2000ApJ...532..170S, 2004AJ....128..569M, 2012MNRAS.425..878M, cai2017, 2020NatAs...4..670A}. Blobs of Ly$\alpha$ are preferentially found in dense regions of star-forming galaxies \citep{2004AJ....128..569M, 2012MNRAS.425..878M}, and typically have sizes of 50-100 kpc (see e.g. Fig.6 of \citealt{cai2017}). In this case, one can well expect to find numerous merger-induced blobs of ionized gas around high-z galaxies, considering that the majority of the \ha\ blobs in our sample are associated mergers, and that galaxy-galaxy interactions/mergers happen more frequently at high redshifts. On the other hand, however, unlike high-z Ly$\alpha$ blobs, the \ha\ blobs in our sample do not prefer dense environment and they are smaller than high-z Ly$\alpha$ blobs. This could be implying that cold-to-warm gas traced by Ly$\alpha$ is more extended than ionized gas, but it could also be due to the limited field of view of MaNGA IFUs. It is thus useful to go outside of the MaNGA IFU in future work, looking for more distant \ha\ emitting regions with larger scales.

%We note that, back in \S~\ref{sec:Deep}, the galaxy-\ha\ blob systems in our sample could be divided into three types according to morphology, which is not necessarily in conflict with the two-group division discussed here, as the latter is based on more physical properties such as metallicity, kinematics, and ionization. In the two-group division, we see that the third type (spot-like \ha\ blobs in isolated late-type galaxies) and some of the second type systems (\ha\ blobs appearing on disks/spirals or tidal structures) constitute this first group, and the first type (\ha\ blobs in interacting early-type galaxy systems) plus the rest of the second type systems make up the second group. Information provided by the spectral analyses help to distinguish the different \ha\ blobs in the second type, for which it is hard to visually discriminate tidal structures and faint disks/spirals.

\subsection{Sample selection bias}
\label{bias}

Since this work is based on a visually selected sample, there is potential bias associated with 
the selection criteria. For instance, \ha\ blobs that are physically offset from the nearby 
galaxy but overlap with the galaxy center along the line of sight could be hard to visually 
identify. This could bias our selected \ha\ blobs to be more concentrated to the main planes of their nearby galaxies, making them more likely to exhibit similar rotational velocity as their nearby galaxies even if they are independent satellites. In addition, \ha\ blobs that lie outside the MaNGA footprints would also be missed 
during the sample selection. As a result, our selected \ha\ blobs all show certain offset from 
their nearby galaxies, but are not too far away from the galaxy centers. This explains why we 
might have a number of star-forming clumps in the outskirt of galaxies (the first group) identified as 
\ha\ blobs. On the other hand, for galaxy systems with two or more members, MaNGA's footprints 
often cover much larger distances. Thus, we could in principle find more distant \ha\ blobs using MaNGA in interacting systems. Moreover, even within the MaNGA sample, visual inspection can lead to different results (see Appendix~\ref{bait}). Our sample includes \ha\ blobs not selected by \cite{Bait-Wadadekar-Barway-19}, but we also missed a few of \cite{Bait-Wadadekar-Barway-19}'s objects purely due to different visual selection criteria. Therefore, both samples are likely biased to some degree. A truly comprehensive sample might include \ha\ blob-like objects with a variety of physical properties and different origins.
Still it is possible that some physical process would produce observable \ha\ blob-like objects more frequently than others.

Finally, an important question related to the sample selection method is whether it is possible to set a quantitative criterion to define \ha\ blobs, instead of purely relying on the visual inspection. Our visual inspection method would encounter ambiguity when there are several potential candidate \ha\ blobs, like in system 3 and system 4, where there are a few other closed contours in the outskirt. In such cases, we only pick the one that is most significantly detected (i.e. the blob with the highest level of \ha\ contour). Therefore, it would be useful if an additional quantitative cut can be applied. It was difficult to set such a criterion a priori, as the true identity of this kind of objects was in question. With our current sample, one potential solution is to use results like those in \autoref{fig:HaF_EW} to set a cut in $\rm \Sigma_{H\alpha}$ and $\rm EW(H\alpha)$, for \ha\ blobs tend to form a distinct branch compared with the other spaxels in the systems. This numerical criterion, however, would still differ for different systems. For example, if one draws a straight demarcation line in this plane to isolate \ha\ blobs, the intercept cannot be a constant in our current sample. In addition, if \ha\ blobs are truly a type of tidal remnants, their surface brightness in optical can vary over several orders of magnitude \citep{2008ApJ...689..936J}, making it even more difficult to make a universal cut. Another potential solution is to combine the current observations with the simulations, and see if there is any general pattern in the observables of systems similar to our sample, which we will explore in future work.

\section{Summary}
\label{sec:summary}

In this work we have identified a set of 13 off-galaxy \ha\ blobs using 
integral field spectroscopy from the MaNGA MPL-7 sample, and attempted 
to understand the physical origin of these special systems by examining 
their deep images as well as two-dimensitional maps of a variety of 
physical parameters, including gas-phase metallicity, kinematics and 
ionizing sources. 

The 13 off-galaxy \ha\ blobs are selected from 4639 galaxies in
MaNGA, by visually comparing the \ha\ flux map and the synthetic optical images from MaNGA. A region is identified to be an 
\ha\ blob if it presents significant \ha\ emission but with no counterpart 
emission in the optical image. Although deep imaging reveals 
%three 
different types of morphology and environments in our sample, spectral
measuements of gas-phase metallicity, kinematics, and ionization
indicate that the \ha\ blobs can be divided into two groups with distinct 
origins. 

The larger group contains 8 high-mass galaxies ({\tt gal6}$-${\tt gal13}, with $M_\ast\ga 10^{10}M_\odot$), 
all of which show signatures of galaxy-galaxy interactions. Most of them
are red and retired from star formation, with only the locations of \ha\ blobs forming stars at relatively low rates. These \ha\ blobs are likely products 
of tidal interactions, and used to be part of their nearby galaxies. AGN or 
shock ionization is not detected from our data. Diffuse ionized gas (DIG) is common in these galaxies, and could also contribute to the \ha\ emission of the \ha\ blobs. 
The other group contains 5 low-mass blue galaxy systems 
({\tt gal1}$-${\tt gal5}, with $M_\ast\lesssim10^{10}M_\odot$ and $NUV-r<4$), showing global star formation
activity and low \ha\ velocity dispersion. The \ha\ blobs in this group 
are likely associated with faint disks ({\tt hab3}), spirals ({\tt hab2, hab5}), or dwarf galaxies ({\tt hab1, hab4}), and have on-going star formation. In particular, there are two \ha\ blobs in this group having spot-like optical counterparts ({\tt hab1, hab4}). One of these \ha\ blobs, {\tt hab4}, exhibits strong emission line spectra and the bluest color and is a potential candidate for blue compact dwarfs.

In both groups, the \ha\ blobs appear to co-rotate with the ionized gas in 
the host galaxies, although in certain cases the \ha\ blob is the only region
with ionized gas. This indicates that the \ha\ blobs are still kinematically 
related with the host galaxy. While most \ha\ blobs do not exhibit very distinct 
metallicities compared with the MZR of the local SF regions, their host galaxies 
sometimes show lower metallicities with a flat or even negative radial gradient, making the \ha\ blobs appearing chemically enriched. This could be a result of tidal interactions and/or recent gas inflows.

To have a better understanding of \ha\ blobs, optical imaging with deeper detection limit is needed to unearth underlying components in these regions. It would also be useful to perform muti-component fitting of stellar profiles to make a cleaner separation of \ha\ blobs from their hosts.
With enough data quality, one might be able to isolate different kinematic components associated with \ha\ blobs and their hosts as well. Finally, a better understanding of the ionizing sources of \ha\ blobs, and particularly the contamination from DIG would help to determine their origin, and reduce the uncertainties in the derived chemical abundance. With these observational evidence, we could look for regions with similar properties in sophisticated simulations in future.

\section*{Acknowledgements}
This work is supported by the National Key R\&D Program of China
(grant No. 2018YFA0404502), and the National Science Foundation of
China (grant Nos. 11821303, 11973030, 11761131004, 11761141012, 11890691, 12120101003). RY acknowledges support by NSF AST-1715898 and NASA grant 80NSSC20K0436 subaward S000353.
RAR acknowledges financial support from CNPq (302280/2019-7) and FAPERGS (17/2551-0001144-9).

Funding for the Sloan Digital Sky Survey IV has been provided by the Alfred P. Sloan Foundation, the U.S. Department of Energy Office of Science, and the Participating Institutions. SDSS-IV acknowledges
support and resources from the Center for High-Performance Computing at
the University of Utah. The SDSS web site is www.sdss.org.

SDSS-IV is managed by the Astrophysical Research Consortium for the 
Participating Institutions of the SDSS Collaboration including the 
Brazilian Participation Group, the Carnegie Institution for Science, 
Carnegie Mellon University, the Chilean Participation Group, the French Participation Group, Harvard-Smithsonian Center for Astrophysics, 
Instituto de Astrof\'isica de Canarias, The Johns Hopkins University, Kavli Institute for the Physics and Mathematics of the Universe (IPMU) / 
University of Tokyo, the Korean Participation Group, Lawrence Berkeley National Laboratory, 
Leibniz Institut f\"ur Astrophysik Potsdam (AIP),  
Max-Planck-Institut f\"ur Astronomie (MPIA Heidelberg), 
Max-Planck-Institut f\"ur Astrophysik (MPA Garching), 
Max-Planck-Institut f\"ur Extraterrestrische Physik (MPE), 
National Astronomical Observatories of China, New Mexico State University, 
New York University, University of Notre Dame, 
Observat\'ario Nacional / MCTI, The Ohio State University, 
Pennsylvania State University, Shanghai Astronomical Observatory, 
United Kingdom Participation Group,
Universidad Nacional Aut\'onoma de M\'exico, University of Arizona, 
University of Colorado Boulder, University of Oxford, University of Portsmouth, 
University of Utah, University of Virginia, University of Washington, University of Wisconsin, 
Vanderbilt University, and Yale University.

\section*{Data availability}

This study utilizes observational data from the DR15 of SDSS, which is publicly available\footnote{www.sdss.org/dr15/data\_access/}. The analysis data underlying this article, including all the derived quantities of the 13 \ha\ blobs, will be shared on reasonable request to the corresponding author.

%%%%%%%%%%%%%%%%%%%%%%%%%%%%%%%%%%%%%%%%%%%%%%%%%%



\bibliographystyle{mnras}
\bibliography{example} % if your bibtex file is called example.bib


%%%%%%%%%%%%%%%%%%%%%%%%%%%%%%%%%%%%%%%%%%%%%%%%%%

%%%%%%%%%%%%%%%%% APPENDICES %%%%%%%%%%%%%%%%%%%%%

\appendix
\label{appendix}

\section{Bait et al.'s sample}
\label{bait}

\begin{figure*}
    \centering
    \includegraphics[width=0.24\textwidth]{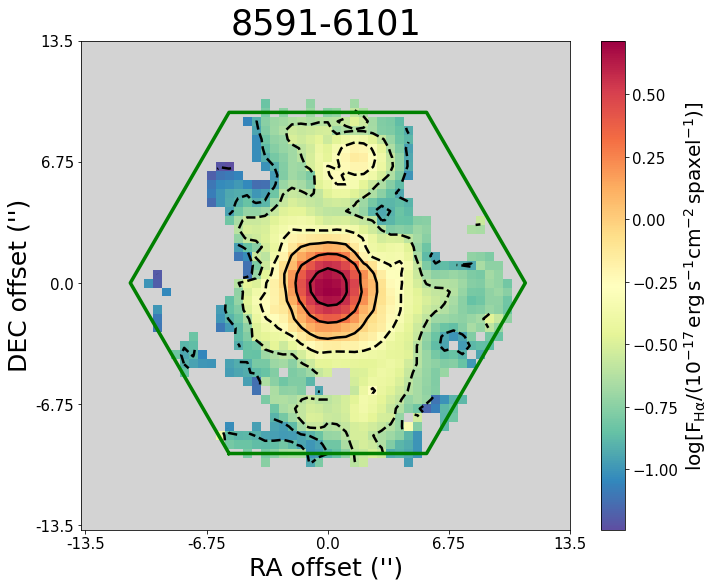}
    \includegraphics[width=0.24\textwidth]{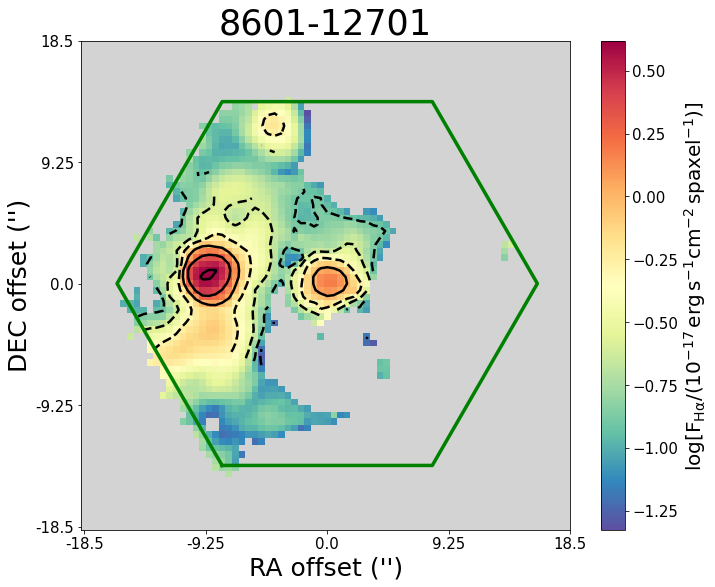}
    \includegraphics[width=0.24\textwidth]{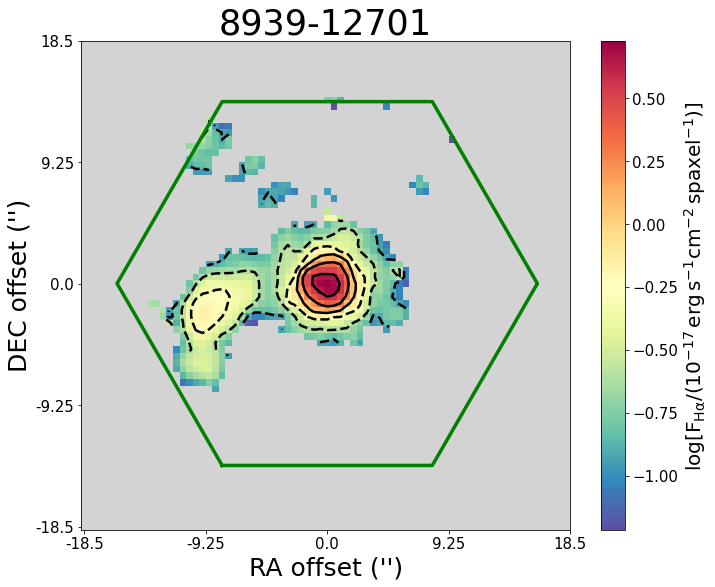}
    \includegraphics[width=0.24\textwidth]{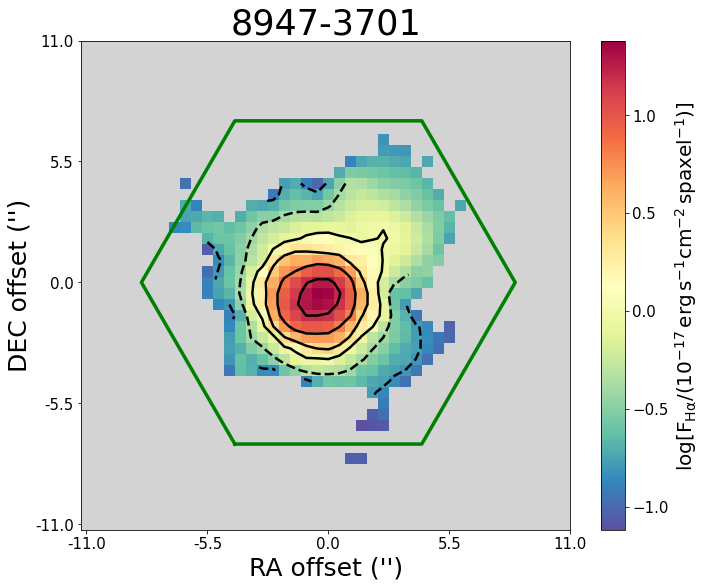}
    \caption{\ha\ intensity maps of Bait et al.'s sample that were not selected by our method. The colored maps are generated from DAP products by cancelling all quality masks. The contours, on the other hand, are plotted with the quality masks on.}
    \label{fig:bait_ha}
\end{figure*}

\begin{figure*}
    \centering
    \includegraphics[width=0.24\textwidth]{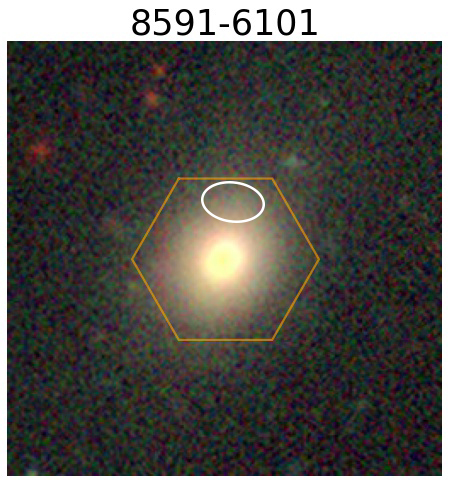}
    \includegraphics[width=0.24\textwidth]{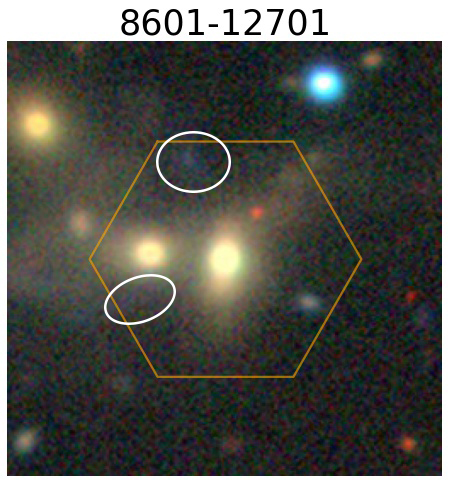}
    \includegraphics[width=0.24\textwidth]{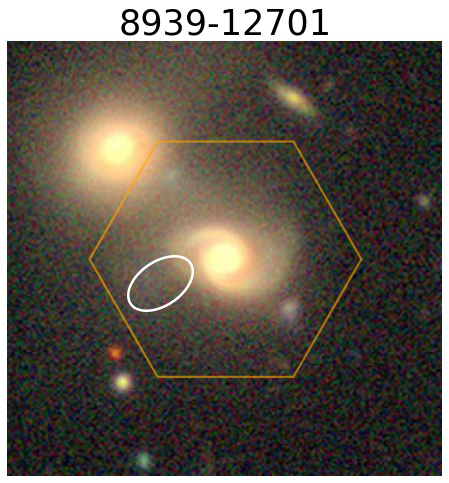}
    \includegraphics[width=0.24\textwidth]{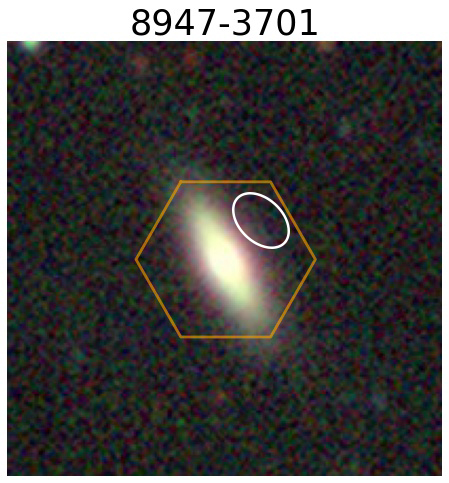}
    \caption{$grz$ color-composite images of Bait et al.'s sample that were not selected by our method. The images are drawn from the Legacy Surveys. The locations of \ha\ blobs are marked with white open ellipses.}
    \label{fig:bait_ls}
\end{figure*}

Back in \S~\ref{sec:Selection}, we mentioned why the \ha\ blob sample in \cite{Bait-Wadadekar-Barway-19} has a poor overlap with ours. In this Appendix, we further discuss the reason as well as the physical properties of the \ha\ blobs in Bait et al.'s sample. To be consistent with Bait et al., we refer to the sample galaxies according to their MaNGA plate and IFU numbers.

\autoref{fig:bait_ha} shows the DAP maps of the Bait et al.'s sample. We note that in system 8601-12701 and 8947-3701, the \ha\ blobs are partially masked. Although the \ha\ blob in the north of system 8601-12701 still has remaining spaxels, the small outer contour makes it appear like noise or artifact and missed by our visual inspection. The DR15 version of DAP masks these spaxels as the binned spectra they belong to do not have reliable stellar kinematic measurements \citep[see][for more details about the bin deconstruction method adopted by the DAP for spectral fitting]{2019AJ....158..231W}. We note that after DR15, MaNGA DAP no longer masks spaxels based on this criterion. In galaxy 8947-3701, if we unmask all spaxels, the `\ha\ blob' does not appear like an isolated blob and does not show closed \ha\ flux contours. The \ha\ blob appear as a half sphere extending from the center of the edge-on galaxy. The locations of its outer contours do not have any detectable optical counterpart. This `\ha\ blob' also shows distinct gas velocity.
Its gas velocity dispersion is relatively high near the edges ($\sim 100$ km/s), but low in the core ($\sim 60$ km/s).
Considering it has both SF- and Seyfert-like ionization, it could be outflowing gas under the influence of both the young massive stars and the central AGN.

On the other hand, the deep image of galaxy 8601-12701 shows a blue and diffuse optical counterpart for the \ha\ blob on the north, as can be seen in \autoref{fig:bait_ls}. Although \cite{Bait-Wadadekar-Barway-19} used the same optical data from the Legacy Surveys, their displayed images do not have enough contrast for the identification of any optical counterpart. The BPT classification of this \ha\ blob is a mix of SF regions and composite regions. Despite its blue color, its emission lines are not very strong (peak EW(\hb )$\sim$10\AA ). The system shows very clear interacting features and the host galaxies exhibit ionization of Seyferts, which can be triggered by the merger \citep{hernquist1989}. Using the H\,{\sc ii} regions in this \ha\ blob, we calculate its metallicity to be about 8.4 through the O3N2 method. This value is similar to what we found in {\tt hab6} and {\tt hab7}.
The estimated stellar mass surface density of this region through SED fitting is very low, which is on the order of $\rm 1.3~M_{\odot }pc^{-2}$ (which is roughly 0.1 on logarithmic scale). This \ha\ blob is thus well above the extrapolated MZR at the low stellar mass surface density end (see \autoref{fig:Global_MZR_ZD}),
which is indicative of a tidal origin, similar to \ha\ blobs in the second group of our sample. 
The `\ha\ blob' (or extended \ha\ emission-line region) in the south-east is likely linked to the tidal features of the galaxy in the east. Since it is closer to the center of the Seyfert galaxy, it exhibits a more AGN-like ionization.
Both \ha\ blobs in this system are corotating with the two galaxies and have low gas velocity dispersion.

The remaining two \ha\ blobs in system 8591-6101 and 8939-12701 were also missed by our sample selection, which is mainly due to the different criteria of our visual inspection procedure. For system 8591-6101, it is hard to tell whether the \ha\ emission is directly from the main body of the galaxy. The BPT diagnostics of this object are composite and LI(N)ER. Use the only two H\,{\sc ii} region spaxels inside the blob, we estimate its metallicity to be roughly 8.4.
We estimate the stellar mass surface density at the location of this \ha\ blob to be roughly $\rm 14~M_{\odot}pc^{-2}$ (which is roughly 1.2 on logarithmic scale). Its MZR thus does not significantly deviate from the median trend. But as we have mentioned before, considering the star light contamination from the central galaxy, the actual optical counterpart of the blob could have a much lower stellar mass surface density.
The host galaxy is classified as LI(N)ER, and the overall \ha\ equivalent width of the system is low, with EW(\ha ) $\lesssim$ 6\AA . The ionization is thus likely dominated by DIG or weak AGN. Its velocity is very different from its nearby regions, and its velocity dispersion is also much lower than the surrounding area. \cite{Bait-Wadadekar-Barway-19} commented that this could be the result of a recent gas-rich merger.

For system 8939-12701, the \ha\ blob appears to be related to one of the spiral arms of the host galaxy. But a closer look on \autoref{fig:bait_ls} reveals that it lies beyond the reach of the apparent northern arm. The host galaxy is again in a merging system. The \ha\ blob is possibly related to the very faint tidal shell to the east and could used to be part of the central galaxy. Interestingly, the host galaxy in system 8939-12701 shows composite- and LI(N)ER-like ionization, but the \ha\ blob shows LI(N)ER- and Seyfert-like ionization. In addition, only the \ha\ blob spaxels have EW(\ha ) > 6\AA . Unlike what we see in many other systems, the ionization and relevant line ratios in 8939-12701 increase from the center to the outskirt. It is possible that the emission line spectra of the central galaxy is contaminated by DIG or very weak SF activities. Or there could be contribution from shocks to the ionization of the \ha\ blob, which requires higher spectral resolution to tell.
The spectra from the nucleus of the central galaxies show double-peak features near \ha , which imply ongoing outflows or accretions. Therefore, it is also possible that the \ha\ blob is created by the AGN driven outflow. This AGN activity might have been triggered by the merger with the galaxy to the northeast.

We can try grouping these \ha\ blobs like we did in the main text. First of all, galaxy 8591-6101, 8601-12701, and 8939-12701 are all massive galaxies. While galaxy 8591-6101 lies in the red sequence, the other two are green-valley galaxies. Galaxy 8947-12701, on the other hand, is a green valley galaxy with $\rm log(M_* /M_{\odot }) = 10.1$. Despite having different ionization, \ha\ blobs in 8601-12701 and 8939-12701 are likely related to tidal interactions and share some similarities of those \ha\ blobs in the second group of our sample. AGN driven outflow is another potential mechanism for creating \ha\ blobs in 8939-12701 and 8947-3701, which is not found in our sample. The origin of the \ha\ blob in 8591-6101 is less clear. The host galaxy is dominated by DIG, similar to systems in the second group. But no clear sign of major interaction is found. The peculiar kinematic feature of the \ha\ blob could be a result of a minor merger instead.

\section{\ha\ intensity maps}
\label{ha_flux}

\begin{figure*}
	\includegraphics[width=0.95\textwidth]{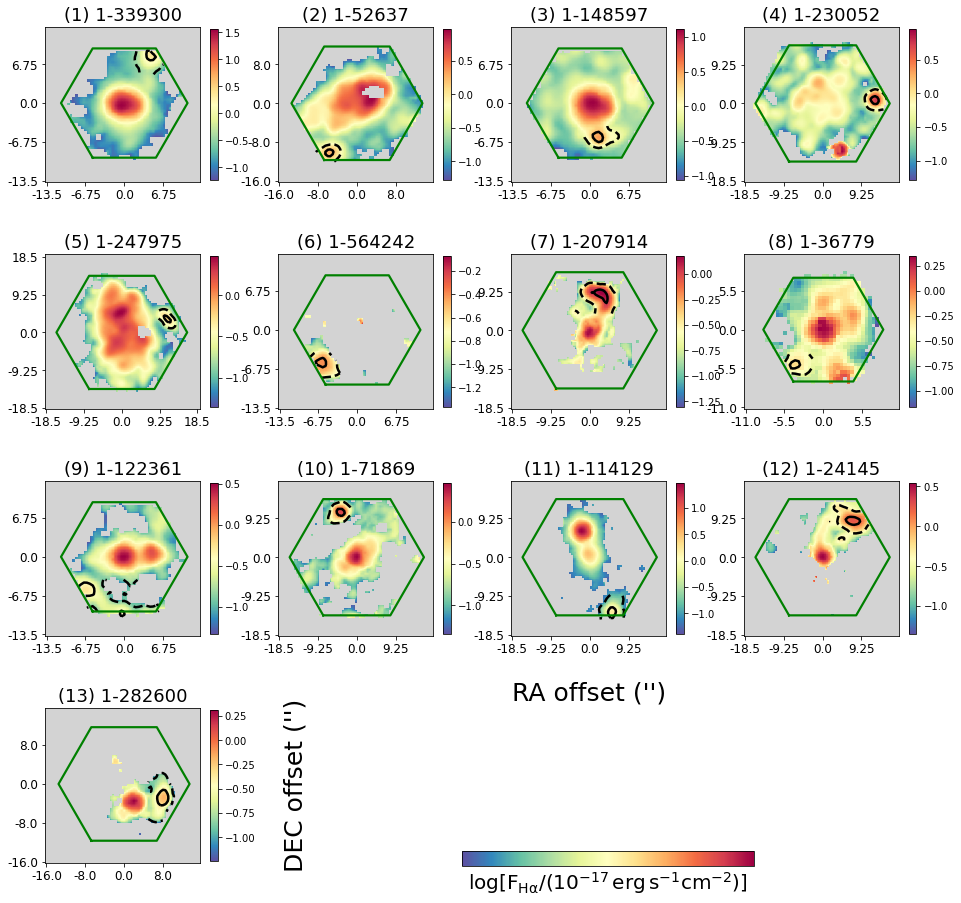}
    \caption{\ha\ intensity maps of our sample galaxies. The green hexagons represent the MaNGA footprints. The dashed and solid contours trace the $\rm 63^{rd}$ and the $\rm 95^{th}$ percentiles of the intensity distribution inside the \ha\ blobs.}
    \label{fig:ha_flux}
\end{figure*}

\autoref{fig:ha_flux} shows the \ha\ intensity maps of our sample galaxies. Compared with the contour plots, these maps better reflect the peaks and valleys of the spatial distribution of the \ha\ intensity. We note that he concentrated \ha\ emission in the south of system 4 comes from a foreground star. The \ha\ emission in the southwest of system 8 is from another galaxy partly enclosed by the hexagonal footprint.

\section{Model-subtracted images}
\label{model_sub}

\begin{figure*}
	\includegraphics[width=0.95\textwidth]{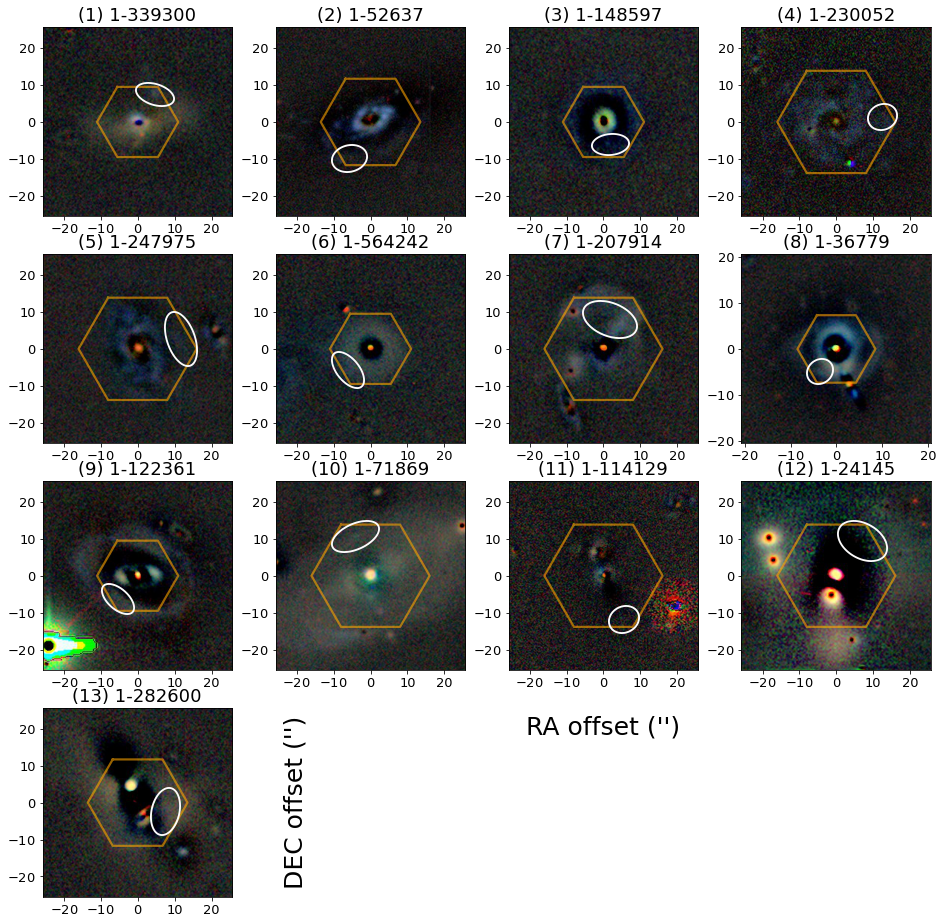}
    \caption{$grz$ color composite images of the residual maps from the Legacy Surveys. The orange hexagons represent the MaNGA footprints. The white ellipses mark the locations of the \ha\ blobs. Note that the optical counterparts of \ha\ blobs in system \redtxt{4} and system \redtxt{11} were identified as independent components by the {\it legacypipe}, and thus were subtracted from the final images.}
    \label{fig:deep_resic}
\end{figure*}

\autoref{fig:deep_resic} shows the optical residual maps from the Legacy Surveys \citep{2019AJ....157..168D}. Identification and subtraction of different sources are done by {\it The Tractor} \citep{2016ascl.soft04008L}. We can see that some underlying faint optical structures (rings, tails, spirals, disks, etc.) become more visible in the residual images, similar to what we have seen in the contrast enhanced images in \S~\ref{sec:Deep}. The spot-like optical counterparts in system 4 and system 11 are identified and subtracted by {\it The Tractor}. 

\section{H {\sc I} observations}
\label{hi}

We obtain H\,{\sc i} observations from HI-MaNGA, which is a follow-up program aiming to collect 21 cm data for MaNGA targets \citep{2019MNRAS.488.3396M}. HI-MaNGA uses the Robert C. Bryd Green Bank Telescope (GBT) and provides single-dish 21 cm data with a FWHM of $8.8^{\prime}$. We find HI-MaNGA observations for 
five of our sample galaxy systems, including system 1, 2, 3, 5, and 6.
We also cross-check our data with the Arecibo Legacy Fast Arecibo L-band Feed Array (ALFALFA) survey \citep{haynes2018}. While ALFALFA seems to have observed the system 4, we find that the actual center of this observation has large offset 
from this galaxy system.

\begin{figure*}
	\includegraphics[width=0.8\textwidth]{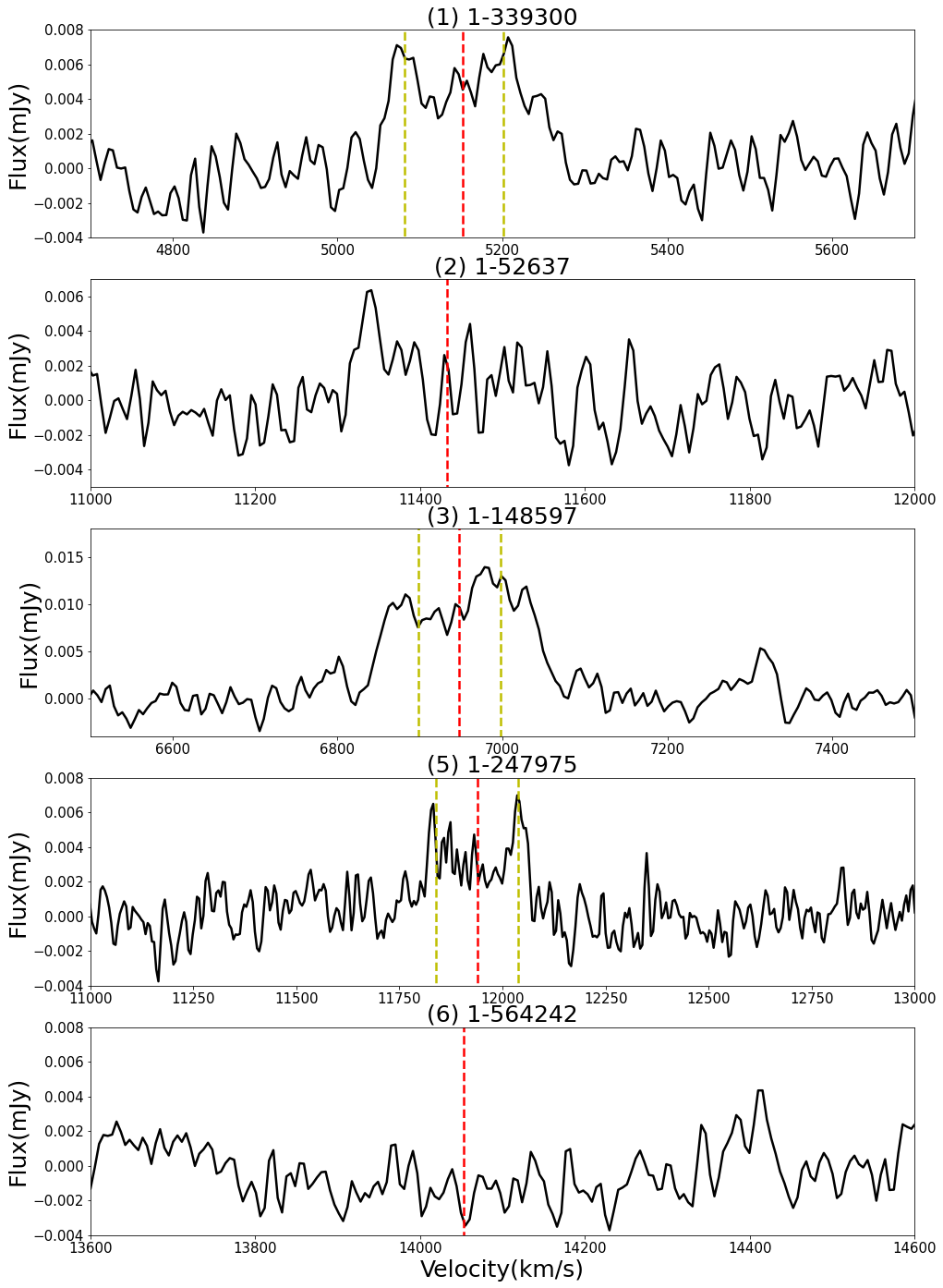}
    \caption{H\,{\sc i} spectra of 5 galaxy system in our sample. The red dashed lines mark the systemic velocity of each system given by the optical spectra. And the yellow dashed lines mark the range of rotational velocity derived by DAP using the ionized gas.}
    \label{fig:a3}
\end{figure*}

Fig.~\ref{fig:a3} shows their H\,{\sc i} spectra along the `velocity' 
(or $cz$) axis. Typical noise level in the flux measurement 
ranges from 0.001 $\sim$ 0.002 mJy. The signal level in system
2 as well as in system 6 is too low to be considered as a valid
detection. The double-peaked profiles indicate the rotation of 
gas inside the galaxy, which in our case perfectly match the 
rotation velocity of ionized gas. One can further estimate the 
total neutral gas content with integrated H\,{\sc i} intensity. The H\,{\sc i} 
masses for these observations are:
log(M$\rm _{HI}$/M$_{\odot}$) = 9.13 for system 1;
log(M$\rm _{HI}$/M$_{\odot}$) = 9.44 for system 2;
log(M$\rm _{HI}$/M$_{\odot}$) = 9.70 for system 3;
log(M$\rm _{HI}$/M$_{\odot}$)=9.84 for system 5;
log(M$\rm _{HI}$/M$_{\odot}$) < 9.47 for system 6.

In system 2, system 5, and system 6, there appear to be very weak off-galaxy H\,{\sc i} emission in the spectra. 
However, their signals are comparable to the noise 
level. In addition, there are discrepancies between their line 
centers and the ionized gas velocity of \ha\ blobs. The \ha\ blob in system 6 is redshifted by $300\sim400$ km/s relative to the systemic velocity of the galaxy which is at 14,053km/s, making it very consistent with the H\,{\sc i} line at 14,400 km/s. Also, the 
central galaxy is line-less. Therefore, this \ha\ blob could
potentially be associated with a cloud of neutral gas indicated by the H\,{\sc i} signal. Regardless, 
spectrum with higher S/N would be helpful for verification.

%%%%%%%%%%%%%%%%%%%%%%%%%%%%%%%%%%%%%%%%%%%%%%%%%%

% Don't change these lines
\bsp	% typesetting comment
\label{lastpage}
\end{document}